\documentclass[preprint,review, 12pt, authoryear]{elsarticle}

\usepackage{amssymb}

\usepackage{amsthm}
\usepackage{amsmath}
\usepackage{marginnote}

\usepackage[left=1in, right=1in]{geometry}

\usepackage{xcolor}
\usepackage{soul}
\usepackage{lineno}

\newcommand{\hl}[1]{{\color{black}{#1}}}

\DeclareMathOperator{\sinc}{sinc}

\journal{Ocean Modelling}

\begin{document}

\begin{frontmatter}



\title{The submesoscale, the finescale and their interaction at a mixed layer front}


\author[a]{Vicky Verma}
\author[a]{Hieu T. Pham}
\author[a,b]{Sutanu Sarkar}

\address[a]{Dept. of Mechanical and Aerospace Engineering, University of California, San Diego}
\address[b]{Scripps Institute of Oceanography, University of California, San Diego}
              
\begin{abstract}
The spindown of a geostrophically balanced density front in an upper-ocean mixed layer is simulated with a large eddy simulation (LES) model that resolves O(1000) m down to O(1) m scale. Our goal is to examine the interaction between the submesoscale and the turbulent finescale, and another related goal is to use the turbulence-resolving simulation to better characterize vertical transport, frontogenesis and dissipative processes. The flow passes through symmetric and baroclinic instabilities, spawns vortex filaments of O (100) m thickness as well as larger eddies with cross-front velocity as large as the along-front velocity, and develops turbulence that  is spatially localized and organized. A O(100) m physical-space filter is applied to the simulated flow so that the coherent  submesoscale is separated from the finescale in a decomposition that preserves the spatial organization of the flow unlike the typical practice of a split into a frontal average and a fluctuation that obscures the coherent submesoscale. The energy spectrum exhibits a change of slope at O(100) m with an approximately -5/3 slope over a subrange of the finescale. Analysis of the submesoscale vertical velocity (as large as 5 mm/s) reveals that downwelling is limited to the thin vortex filaments and the junction of the submesoscale eddies with these filaments while upwelling occurs over spatially extensive regions in the eddies. Conditional averaging shows that heavier (lighter) fluid is preferentially downwelled (upwelled) by these coherent submesoscale structures leading to an overall buoyancy flux that is restratifying. The submesoscale is unbalanced with local Rossby number as large as 5. The kinetic energy (KE) transport equations are evaluated separately for the submesoscale and the finescale to understand energy pathways in this problem. The buoyancy flux (associated with coherent motions) transfers the potential energy of the front and acts as the primary source of submesoscale KE which is then transported across the front with a fraction transferred to the finescale. The transfer, limited to thin regions of O(100) m horizontal width, is accomplished by primarily horizontal strain in the upper 10 m and by vertical shear in the rest of the 50 m deep mixed layer. Frontogenetic mechanisms are diagnosed through analysis of the transport equation for squared buoyancy gradient. Horizontal strain is the primary frontogenetic term that is especially strong in the near-surface layer. The frontogenesis is counteracted primarily by horizontal diffusion in the top 10 m while, further below, the balance is with the horizontal gradient of vertical velocity.

\end{abstract}


\end{frontmatter}

\section{Introduction} \label{intro} 

The mixed layer in the upper ocean contains  fronts (regions that have sharp density gradient) that exhibit lateral density variability at sub-10 km scale, e.g. \citet{HosegoodGM:2006,TimmermansCT:2012,SenguptaRRSP:2016}. Submesoscale (0.1-10 km) instabilities (e.g. \citet{ThomasTM:2008,McWilliams:2016}) at frontal regions  have the following  important consequences for the state of the upper ocean. There is restratification that changes the local properties of the mixed layer and influences the upper-ocean fluxes that are transmitted through the base of the mixed layer. The  vertical velocity is enhanced, thereby promoting the exchange of heat, nutrients and material across the surface layer. There are unbalanced motions  with O(1) values of Rossby number ($Ro_l = u/fl $ where $u$ is a characteristic horizontal velocity, $l$ is a characteristic horizontal length scale, and $f$ the Coriolis parameter) that link the rotationally controlled mesoscale to three-dimensional turbulent motions.

The stratified, rotating flow at a  front  is susceptible to various  spontaneous, unforced  instabilities  whose nonlinear evolution leads to multiscale dynamics which will be studied here. The canonical Eady model \citep{Eady:1949} of an inviscid  geostrophically balanced  flow with uniform values for  lateral buoyancy gradient ($M^2 = |\partial b/\partial y|$), buoyancy frequency ($N$), and Coriolis parameter ($f$)  is governed by two nondimensional parameters, Richardson number ($Ri = N^2/S^2$ with $N$ the buoyancy frequency and $S = M^2/f$ the vertical shear) and the lateral buoyancy parameter ($M^2/f^2$).  The linear evolution is dominated by symmetric instability (SI) when Richardson number lies between 0.25 and 0.95, and by  baroclinic instability (BI) when $Ri > 0.95$ as shown by \citet{Stone:1966}. The  SI mode has no along-front variability  (wavenumber $k =0$)  and the BI mode has no cross-front variability (wavenumber $l=0$). The slantwise currents that develop  during SI have vertical shear which becomes unstable to the Kelvin-Helmholtz (KH) mode as demonstrated in the 2-D simulations of \citet{TaylorF:2009} while  3-D simulations \citep{AroboneS:2015}  exhibit additional mixed modes ($k\neq0$, $l\neq0$) before the KH instability develops. BI has received much attention owing to its relation to the growth of deep mesoscale eddies in the strongly stratified interior of the ocean. The weaker stratification of the mixed layer allows smaller-scale, shallower submesoscale eddies to develop   through a version of the BI, sometimes called  the  mixed layer instability (MLI), that leads to a potent restratifying buoyancy flux (e.g. \citet{BoccalettiFF:2007,FoxKemperFH:2008}). In cases with initial $Ri$ between 0.25 and 0.95, SI develops preferentially during the initial evolution, but eventually $Ri$ increases to exceed 0.95  so that BI dominates (e.g \citet{HaineM:1998,FoxKemperFH:2008}).

An important aspect of frontal evolution is frontogenesis \citep{HoskinsB:1972,Hoskins:1982}, a process by which the width of a front decreases and  there is an increase in horizontal buoyancy gradient, vertical velocity and vertical vorticity. Frontogenesis is particularly strong near the surface \citep{LapeyreKH:2006}, and \citet{Spall:1995} suggests that large vertical velocity needed to subduct a fluid parcel to the bottom of the mixed layer can occur through frontogenesis. Simulations of model fronts (e.g. \citet{MahadevanT:2006,BoccalettiFF:2007,FoxKemperFH:2008}) show that  submesoscale processes in the surface layer are indeed effective in restratifying the front as is also found in regional-scale models (e.g. \citet{CapetMMS:2008a,CapetMMS:2008b}) that resolve the mesoscale-submesoscale transition. Nonhydrostatic, high-resolution numerical models such as that  in the present work are able to access the progression of frontal thinning to the scale of turbulent eddies  where frontogenesis is arrested. At these scales, the dynamics is 
better described by turbulent thermal wind (TTW) balance rather than geostrophic balance \citep{GulaMM:2014, McWilliamsGMRS:2015, McWilliams:2016, WenegratM:2016, McWilliams:2017, SullivanM:2018}. 

\hl{Our understanding of dynamics at the submesoscale, that had progressed primarily by  theoretical models and numerical experiments, has recently benefited from Lagrangian-based observations. \citet{DAsaro_etal:2018} in a surface-drifter study in the northern Gulf of Mexico (DeSoto canyon region) found that some of the drifters cluster in a long, thin sharpened front which then rolls up into a cyclonic eddy which is only few kilometers in diameter. The drifters also reveal {\em zipper} structures where two sharpened fronts merge into one that wraps around the eddy. The front and eddy are convergent and have large positive vertical vorticity with values that are particularly strong in the convergent zipper region. Measurement of vertical velocity by floats reveals downwelling in the convergent region with magnitude  as large as 1-2 $\rm{cm \, s^{-1}}$. Previously, in another observational study performed in the northern Gulf of Mexico using a large number of surface drifters, \citet{Poje_etal:2014} found energetic submesoscale turbulence that had considerable effect on the local dispersion in the submesoscale range. In a novel two-point, synchronized measurement of velocity on two parallel tracks, \citet{ShcherbinaDLKMM:2013} found that the overall structure of the vertical vorticity on a horizontal plane is that of strands of strong cyclonic vorticity merged in the background of weak anticyclonic vorticity. Moreover, large positive vertical vorticity was found to be correlated with  large strain rate.}

In the present work, we explore submesoscale/finescale dynamics during the evolution of SI and BI in an unforced front. Our tool is  high-resolution, nonhydrostatic modeling. Forcing by wind, buoyancy and waves in frontal regions lead to additional processes that, although of interest, are not studied here. Baroclinic instabilities in the mixed layer are able to develop in the presence of wind, surface waves and convection (e.g. \citet{MahadevanTF:2010, HamlingtonRF:2014, CalliesF:2018}) but their overall importance to  vertical exchange, lateral stirring and mixing depends on the strength of the forcing.  
  
The increasing  resolution and accuracy of recent 3-D non-hydrostatic numerical models has enabled access to dynamics at the smallest submesoscales as well as the finescale turbulence as summarized below for unforced frontal problems. \citet{SkyllingstadS:2012} simulated the evolution of BI in a warm filament (double-front configuration with lateral and along-front periodicity) that has $\Delta T = 0.08$ K  across a front width $L =1.2$ km  in a $H = 80$ m deep mixed layer with an LES model that was initially run at 6 m resolution with 8 hrs of surface cooling, and then continued without forcing after interpolation on to a 3 m grid. In fronts with finite width, $L$, in addition to $M_0^2/f^2$ ($M_0^2 = |\Delta b/L|$, where $\Delta b$ is the buoyancy change across the front width $L$) and $Ri$ that govern the Eady problem \footnote{The background lateral buoyancy gradient, $M$, is held constant in the Eady problem while it evolves from it initial value, $M_0$, in a finite-width front}, there is another independent parameter, the Rossby number ($Ro = U_o/fL = M_0^2 H/f^2L$) that is introduced by the initial horizontal shear. The dimensional  parameters given in \citet{SkyllingstadS:2012} lead to $M_0^2/f^2 \approx 6.67$, $Ro \approx 0.44$.  The initially unstratified surface layer develops stratification with $Ri \approx 2$ after the surface cooling. A major result of \citet{SkyllingstadS:2012} is that turbulence develops at isolated small-scale features on the sharpening baroclinic wave instead of the classical picture of a continuous energy cascade through intermediate wavenumbers. Further analysis~\citep{SamelsonS:2016} of the model results showed that frontogenesis  leads to features with unstable vertical shear ($Ri < 0.25$) where turbulence is found. \citet{OzgokmenPFH:2011} conducted LES of BI in a front that is 3 km wide and has a 80 m deep mixed layer with a focus on drifter-based sampling strategies. They considered a  $Ro = 0.066$ front with  strong rotational control and dynamics different from our present interest. \citet{AroboneS:2015} simulated the nonlinear evolution of SI into turbulence with $Ri = 0.5$ and $M^2/f^2 = 16$ on a grid with $1024 \times 1024 \times 256$ points and, in addition to the 2-D (in lateral-vertical plane) KH instability found by \citet{TaylorF:2009}, found a tertiary instability (in the downfront-vertical plane) that preceded 3-D turbulence. The integration time and domain size were not sufficiently large to allow BI to form. \citet{StamperT:2017} performed simulations of the Eady problem with  various $Ri$ between 0.25 and 1, and fixed $M^2/f^2 = 10$ on a grid  with 2.4 m horizontal resolution and 0.8 m vertical resolution.  They found that mixed modes ($k\neq 0$ and $l \neq0$) contain the maximum turbulent kinetic energy (TKE) in the transition period between SI and BI (dominates later when $Ri$ exceeds 0.95), and the time-integrated buoyancy term dominates the shear production term in all cases. The Reynolds number, $Re = U_0 H/\nu = M_0^2 H^2    /f \nu $  is another independent parameter in the simulations and takes the value of $Re = 1.6\times10^4$ in \citet{OzgokmenPFH:2011}, $Re = 8 \times 10^4$ in the 3-D turbulence case of \citet{AroboneS:2015}, $Re = 4.1 \times 10^4$ in \citet{StamperT:2017}, and is unspecified in \citet{SkyllingstadS:2012}.

In the present work, we conduct high-resolution  LES of  a front at moderate values of  $Ri = 0.26$ and $Ro = 0.32$, a relatively high $Re = 2.67 \times 10^6$, and a  2 m isotropic grid in a $4098 \, \rm{m} \times 6146 \, \rm{m} \times 130 \, \rm{m}$ domain. The preceding paragraph shows that our knowledge of submesoscale and turbulence dynamics in unforced fronts has progressed owing to the deployment of non-hydrostatic, 3-D numerical models. However, the  relative contributions of the submesoscale and the finescale turbulence to vertical velocity, vorticity, dissipation and mixing requires further quantification  and the interactions between the submesoscale and the turbulence require improved understanding. In particular, since turbulence is spatially localized by the coherent submesoscale structures, a simple decomposition of the flow between a mean (e.g., along-front averaging) and a fluctuation is inadequate to understand the interaction between the coherent structures that emerge and the finescale turbulence. Instead of frontal averaging, we apply an explicit spatial filter to the simulation data that, as will be demonstrated, effectively separates the small-scale, resolved turbulence (the finescale) from the coherent submesoscale structures,  and helps characterize their different roles in vertical transport, frontogenesis/frontolysis, dissipation and mixing. The submesoscale-finescale decomposition also enables separation of the  balance equations between  the individual KE components, and a  quantification (local in physical space and nonlocal in wavenumber space) of the submesoscale-turbulence energy transfer that is better suited to the type of turbulence that develops in this problem. 
 
The paper is structured as follows. The setup of a model front and the numerical model is presented in Section~\ref{sec:setup}. Section~\ref{sec:spatial_filtering} describes the spatial filtering scheme used for separating the submesoscale from the finescale motions. The frontal evolution under symmetric and baroclinic instabilities is summarized in Section~\ref{sec:frontal_instabilities}. The choice of the length scale for the submesoscale-finescale decomposition is also explained. Section~\ref{sec:scale_separation} describes the properties of the submesoscale and finescale fields with a focus on the   vertical velocity and restratifying fluxes.  The kinetic energy in the individual submesoscale and finescale components, and their transport equations  are quantified in  Section~\ref{sec:kinetic_energy} to better understand the energy pathways. Submesoscale frontogenesis by the finite-amplitude baroclinic instability is investigated in Section~\ref{sec:frontogenesis} by diagnosing the balance equation for the horizontal buoyancy gradient. Finally, we conclude in Section~\ref{sec:discussion_and_conclusion} with a summary of the results and discussion in the context of previous work.
\section{Model setup} \label{sec:setup}
The model consists of a density front that is geostrophically balanced by a surface jet. In Cartesian coordinates, the front and the jet align with the x-direction (along-front) and the lateral (cross-front) density variation is in the y-direction. The z-direction (vertical) coincides with the axis of rotation. A linear equation of state, ${\rho}/{\rho_0} = - \alpha T $,  is used to relate the temperature deviation, $T$, from a reference temperature $T_0$ to that of density deviation, $\rho$, from a reference density $\rho_0 = 1028 \, \rm{kg \, m^{-3}}$, where $\alpha = 2 \times 10^{-4} \, \rm{K}^{-1}$ is the coefficient of thermal expansion. 

\begin{figure*}[h!]  
\centering
\includegraphics[clip=true, width=6.0in]{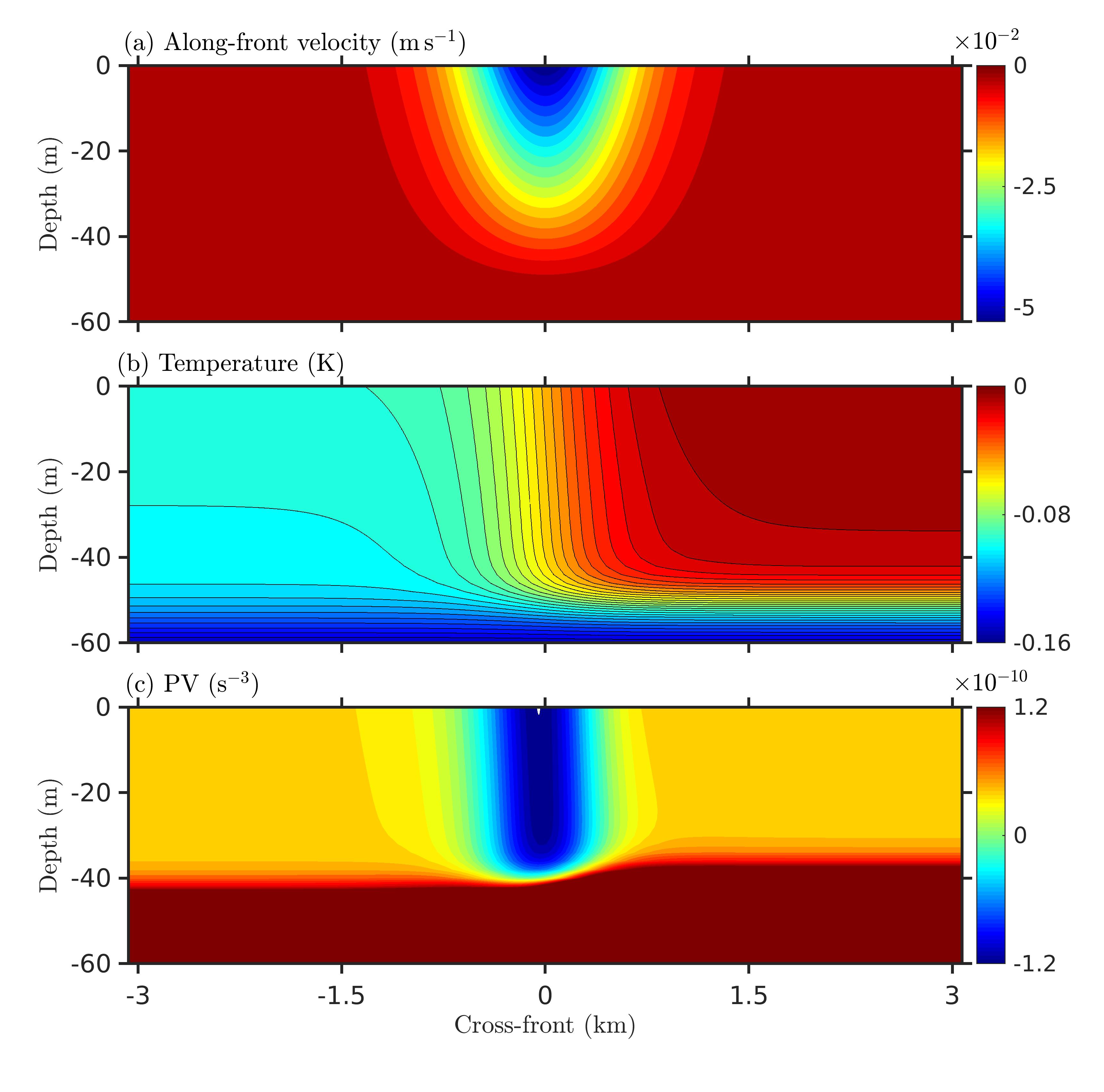}
  \caption{Initial profiles in the model front: (a) along-front velocity, (b) temperature and (c) potential vorticity. The profiles are uniform in the along-front direction and the initial cross-front velocity is zero. The plot of potential vorticity (panel c) shows that the front is unstable to symmetric perturbations.  
 }\label{fig:initial_condition}
\end{figure*}

The front is centered at $y = 0$, has a width of $L =  1.2 \, \rm{km}$ and is confined in a mixed layer with depth $H = 50 \, \rm{m}$, situated over a thermocline. The initial velocity, temperature and potential vorticity vary in the cross-front ($y$) and vertical ($z$) directions as shown in Fig.~\ref{fig:initial_condition}. \hl{The potential vorticity is defined as $\rm{PV} = (\boldsymbol{\omega} + f \mathbf{k})\cdot \mathbf{\nabla}b$, where $\boldsymbol{\omega}$ is the relative vorticity, $\mathbf{k}$ is a unit vector in the vertical direction and $b = \alpha T g$ is the buoyancy.} The temperature variation is given by:
\begin{align} \label{eq:T}
 T(y, z) =  &-\frac{M_0^2 L}{\alpha g}\left\{1-0.25 \left[1+\tanh\left( \frac{y}{0.5 L}\right) \right] 
           \left[1+\tanh\left(\frac{z+H}{\delta_H}\right) \right] \right\}  \nonumber \\
            &+ \frac{0.5}{\alpha g}\left\{\left( N_M^2+N_T^2 \right)z+ \delta_H \left(N_M^2-N_T^2 \right)  \log\left[\frac{\cosh((z+H)/\delta_H)}{\cosh(H/\delta_H)}\right]\right\}.
\end{align}
Here $M_0^2$ is the value of $M^2 = -(g/\rho_0) \partial \rho/ \partial y$ evaluated at the center $y = 0$, and $M^2$ is defined analogous to the square of buoyancy frequency associated with the vertical density gradient, $N^2 = -(g/\rho_0) \partial \rho/\partial z$; the parameters $N^2_M$ and $N^2_T$ are the square of buoyancy frequencies in the mixed layer and the thermocline, respectively; $\delta_H = 5\, \rm{m}$ is a thin region between the mixed layer and the thermocline where the temperature profile joins smoothly from its value in the mixed layer to that in the thermocline;  $\rm{g}  = 9.81 \, \rm{m}  \, \rm{s}^{-2}$ is the gravitational acceleration. In the present simulation, $M^2_0 = 1.5 \times 10^{-7} \, \rm{s^{-2}}$, $N^2_M = 3.0 \times 10^{-7} \, \rm{s^{-2}}$ and $N^2_T = 10^{-5} \, \rm{s^{-2}}$. Thus, the squared buoyancy frequency in the thermocline is two-orders of magnitude larger than that in the mixed layer.

The geostrophic jet, $U(y,z)$, is constructed by integrating the density field to satisfy the thermal wind balance, i.e. $\partial U/\partial z = - M^2/f$, where $f = 1.4 \times 10^{-4} \, \rm{s}^{-1}$ is the Coriolis parameter. Additionally, broadband velocity fluctuations with amplitude of $10^{-4} \, \rm{m} \, \rm{s}^{-1}$ are added to the frontal jet in order to initiate the instabilities. 

The large eddy simulation (LES) approach is used to simulate the model front using non-hydrostatic Navier-Stokes equations under Boussinesq approximation. Along-front velocity $u_1$, cross-front velocity $u_2$, vertical velocity $u_3$, temperature $T$ and dynamic pressure $p$ are advanced in time $t$ as follows:
\begin{align} \label{eq:NS}
 \frac{\partial u_j}{\partial x_j} &= 0,  \nonumber \\
 \frac{\partial u_i}{\partial t} + \frac{\partial u_i u_j}{\partial x_j} + \epsilon_{ijk} f_j u_k &= -\frac{1}{\rho_0} \frac{\partial p}{\partial x_i} + \alpha T g \delta_{i3} + \nu \frac{\partial^2 u_i}{\partial x_j^2} -  \frac{\partial \tau^{sgs}_{ij}} {\partial x_j},   \nonumber \\
 \frac{\partial T}{\partial t} + \frac{\partial u_j T}{\partial x_j} &= \kappa \frac{\partial^2 T}{\partial x_j^2} - \frac{\partial q^{sgs}_j}{\partial x_j}, 
\end{align}
\hl{where i, j, k = 1, 2, 3, and a repeated index implies summation}. Here $\nu$ is the molecular viscosity and  $\kappa$ is the molecular diffusivity;  $\tau^{sgs}_{ij} = - \nu^{sgs} (\partial u_i/\partial x_j + \partial u_j/\partial x_i)$ is the modeled LES subgrid stress tensor and $q^{sgs}_j = -\kappa^{sgs} (\partial T/\partial x_j)$ is the modeled LES subgrid heat flux; and, $\nu^{sgs}$ and $\kappa^{sgs}$ denote the subgrid viscosity and diffusivity, respectively. Parameters $\nu$ and $\kappa$ are related by the Prandtl number, $Pr = \nu/\kappa$; the value of molecular viscosity used is $\nu = 10^{-6} \, \rm{m^2 s^{-1}}$, and the Prandtl number $Pr = 7$. An alternate notation for the velocity components is also used wherein the along-front, cross-front and vertical velocity components are expressed as $u$, $v$ and $w$, respectively.

When Eq.~\ref{eq:NS} is scaled by the velocity scale $U_0 = M_0^2 H/f$, the maximum geostrophic jet velocity at the ocean surface, and the buoyancy scale $N^2_M H$, the {\em non-dimensional} parameters are as follows: the Ekman number, $Ek = \nu/fH^2$, the non-dimensional lateral buoyancy gradient, $M_0^2/f^2$, and the Richardson number, $Ri = N^2_M f^2 / M_0^4$. In the present study, $Ri = 0.26$ and $Ek = 2.86 \times 10^{-6}$. The ratio $M_0^2/f^2 = 7.65 $ is comparable to the values used in the studies of \citet{SkyllingstadS:2012} and \citet{HamlingtonRF:2014}. Also, note that the Rossby number, $Ro = U_0/fL$, based on the initial horizontal shear is 0.32 and the Reynolds number, $Re = U_0 H/\nu$, is $2.67 \times 10^{6}$. 

The subgrid fluxes need parameterization in LES. Following \citet{DucrosCL:1996}, the subgrid viscosity, $\nu^{sgs}$, is computed dynamically at every grid point $(i, j, k)$ using a local velocity structure function $F$:
\begin{equation}
\nu^{sgs} = 0.0014 C_K^{-3/2} \Delta \left[F(x_i, \Delta x_i,t) \right]^{1/2}, \,
\end{equation}
where $C_K = 0.5$ is the Kolmogorov constant, $\Delta = || \Delta x_i ||$ is the magnitude of the filter grid spacing, and
\begin{align}
 F(x,\Delta x_i,t)  = &\frac{1}{4} (||\tilde{\mathbf{u}}_{i+1,j,k} - \tilde{\mathbf{u}}_{i,j,k}||^2 + ||\tilde{\mathbf{u}}_{i-1,j,k} - \tilde{\mathbf{u}}_{i,j,k}||^2 \nonumber \\
                      &+ ||\tilde{\mathbf{u}}_{i,j+1,k} - \tilde{\mathbf{u}}_{i,j,k}||^2 + ||\tilde{\mathbf{u}}_{i,j-1,k} - \tilde{\mathbf{u}}_{i,j,k}||^2). 
\end{align}
Here, $\tilde{\mathbf{u}}_{i,j,k}$ is the three-component velocity field that is obtained after passing the LES velocity through a discrete Laplacian high-pass filter. The model leads to substantial subgrid viscosity only at grid points with large velocity fluctuations and has been used previously in several problems including the oceanic examples of turbulent  baroclinic eddies \citep{SkyllingstadS:2012} and the formation of gravity currents from strong fronts \citep{PhamS:2018}. The subgrid diffusivity for temperature is taken to be equal to the subgrid viscosity, i.e., the subgrid Prandtl number $Pr_{sgs} = 1$.

The computational domain is a rectangular box bounded by $0 \le x \le 4098 \, \rm{m}$,  $-3073 \, \rm{m} \le y \le 3073 \, \rm{m}$ and $-130 \, \rm{m} \le z \le 0 $. A uniform grid with $2050 \times 3074 \times 66$ points provides a grid resolution of 2 m in each direction. The use of a fine grid resolution is needed in order to resolve the 3-D overturning motions associated with the vortex filaments that develop during the nonlinear evolution of BI. The domain size is chosen to accommodate the growth of the most unstable baroclinic mode \citep{Stone:1966} whose wavelength, $L_b$, and the time scale, $\tau_b$, are:
\begin{equation} \label{eq:BI}
L_b  =  2\pi H \frac{M_0^2}{f^2} \sqrt{\frac{1+Ri}{5/2}}, \quad
\tau_b = \sqrt{\frac{54}{5}} \frac{\sqrt{1+Ri}}{f} \, . 
\end{equation}
With the parameters used in the present study, the chosen domain is large enough to accommodate at least two wavelengths of the most unstable baroclinic mode. The simulations are run for about 100 h and are terminated when the coherent submesoscale eddies become comparable to the width of the front.

The boundary condition in the along-front (x) direction is periodic. Free-slip on the velocity and no-flux on the temperature are used as the boundary conditions at the  surface (z=0) and side boundaries. At the bottom boundary, free-slip  is used for the velocity and a constant heat flux corresponding to the temperature gradient in the pycnocline  is imposed. Sponge layers are employed at the lateral and bottom boundaries to prevent reflection of spurious waves. The sponge layers at the lateral boundaries have a thickness of $64 \, \rm{m}$; the sponge layer at the bottom boundary is $20 \, \rm{m}$ thick. The governing equations (Eq.~\ref{eq:NS}) are advanced in time using a mixed third-order Runge-Kutta (for advective fluxes) and Crank-Nicolson (for diffusive fluxes). Second-order finite difference discretization is used to compute spatial derivatives. The dynamic pressure is obtained by solving the Poisson equation with a multi-grid iterative method.

\section{Separation into the submesoscale and the finescale} \label{sec:spatial_filtering}
The nonlinearly evolving frontal instabilities lead to three-dimensional turbulence that is highly local and  concentrated at the coherent structures \citep{SkyllingstadS:2012,StamperT:2017}. In order to understand how the three-dimensional finescale interacts with the large-scale flow, a spatial filter is used to decompose the flow between finescale and submesoscale components in the physical space. The submesoscale is obtained by applying a low-pass spatial filter to the LES field \footnote{This low-pass filter is an additional explicit spatial filter that has a \hl{length scale} that is much larger than the grid length scale which is the filter implicit in the LES field.}. Here, we discuss the filtering method and the transport  equations for the submesoscale and the finescale. The choice of the spatial filter length scale and the dynamics of the submesoscale are discussed later in section~\ref{sec:scale_separation}. 

The  LES field variable ($\phi$) is decomposed into the submesoscale ($\overline{\phi}$)  and the finescale ($\phi^{\prime \prime}$) as follows:
\begin{equation}
\phi =  \overline{\phi} + \phi^{\prime \prime} \, .
\end{equation}
The  submesoscale is obtained by the spatial filtering defined as a convolution:
\begin{equation}
 \overline{\phi}({\mathbf{x}, t}) \equiv G*\phi = \int G(\mathbf{r}) \phi (\mathbf{x}-\mathbf{r}, t) d\mathbf{r},
\end{equation}
where $\phi = (u, v, w, p, T)$ is a LES field variable, $G(\mathbf{r})$ is the filter kernel, and $\mathbf{r}$ is a position vector measured relative to $\mathbf{x}$. 
The  finescale component is obtained by subtracting the submesoscale component from the LES field, $\phi^{\prime \prime} = \phi - \overline{\phi}$. Notice that $(.)^{\prime \prime}$ is used to denote the finescale component and $(.)^{\prime}$ is reserved to denote the fluctuation, $\phi^{\prime} = \phi - \langle \phi \rangle_{x}$, with respect to the Reynolds average where the average is taken to be the along-front average, $\langle \phi \rangle_{x} = (1/L_x)\int_0^{L_x} \phi(\mathbf{x}, t) dx$.

In this study, a two dimensional Lanczos filter has been used in the horizontal plane. The filter kernel in one dimension is given by
\begin{equation}
 G_{1D}(\zeta) = \sinc(k_c \zeta) \sinc \left(\frac{k_c \zeta}{a} \right); \quad \sinc(k_c \zeta) = \frac{\sin(k_c \zeta)}{k_c \zeta}.
\end{equation}
Here, $a$ is a non-zero positive integer, $k_c$ is a cutoff wavenumber, and $\zeta$ is the distance from the position where the filter is applied. The extension of the filter kernel to two dimensions is straight forward and is given by
\begin{equation}
 G_{2D}(\zeta, \eta) = G_{1D}(\zeta)G_{1D}(\eta)\, ,
 \label{eq:2DLanczos}
\end{equation}
where $\zeta$ and $\eta$ are distances in the $x$- and $y$-directions, respectively. The Lanczos filter provides the advantage of a sharp cut-off in the wavenumber space. This filter has been used in previous studies to separate the very-large scale flow (mesoscale) from the  submesoscale \citep{HazaOH:2016,MensaG:2013}.

To  understand the dynamical consequences of the separation into the submesoscale and finescale, it is useful to derive the equations that govern these individual components from the original  Eq.~\ref{eq:NS}. Since the filter function is homogeneous, continuity is satisfied at both  scales, i.e.,
\begin{equation}
 \frac{\partial \bar{u}_j}{\partial x_j} = 0, \quad \frac{\partial u^{\prime \prime}_j}{\partial x_j} = 0.
\end{equation} 
The momentum and temperature transport equations at the submesoscale become
\begin{align} \label{eq:sms_eqn}
 \frac{\partial \overline{u}_i}{\partial t} +
  \frac{\partial \overline{u}_i \overline{u}_j}{\partial x_j} 
  + \epsilon_{ijk} f_j \overline{u}_k &= -\frac{1}{\rho_0} \frac{\partial \overline{p}}{\partial x_i} + \alpha \overline{T}g \delta_{i3} + \nu \frac{\partial^2 \overline{u}_i}{\partial x_j^2} - \frac{\partial \overline{\tau}^{sgs}_{ij}}{\partial x_j} - \frac{\partial \tau^{R}_{ij}}{\partial x_j}, \notag \\
 \frac{\partial \overline{T}}{\partial t} + \frac{\partial \overline{u}_j \overline{T}}{\partial x_j} &= \kappa \frac{\partial^2 \overline{T}}{\partial x_j^2} - \frac{\partial \overline{q}^{sgs}_j}{\partial x_j} - \frac{\partial q^{R}_j}{\partial x_j}, 
\end{align}
where $\tau^R_{ij} = \overline{u_i u_j} -\overline{u}_i \overline{u}_j$ is the residual stress and $q^{R}_j = \overline{u_j T} - \overline{u}_j \overline{T}$ is the residual heat flux; $\overline{{\tau}_{ij}}^{sgs}$ and $\overline{{q_j}}^{sgs}$ are the model subgrid stress and heat flux, respectively, at the submesoscale after applying the additional Lanczos filter. 

At the finescale, the momentum and temperature transport equations become
\begin{align} \label{eq:fs_eqn} 
 \frac{\partial u^{\prime \prime}_i}{\partial t} 
 + \overline{u}_j \frac{\partial  u^{\prime \prime}_i}{\partial x_j}
 +  u^{\prime \prime}_j \frac{\partial \overline{u}_i} {\partial x_j}
  + \frac{\partial u^{\prime \prime}_i u^{\prime \prime}_j}{\partial x_j} 
 + \epsilon_{ijk} f_j u^{\prime \prime}_k  &= -\frac{1}{\rho_0} \frac{\partial p^{\prime \prime}}{\partial x_i} + \alpha T^{\prime \prime} g \delta_{i3} + \nu \frac{\partial^2 u^{\prime \prime}_i}{\partial x_j^2}
-  \frac{{\partial \tau_{ij}^{\prime \prime \, sgs}}}  {\partial x_j}
 + \frac{\partial \tau^{R}_{ij}}{\partial x_j}, 
\notag \\
 \frac{\partial T^{\prime \prime}}{\partial t} 
+ \overline{u}_j \frac{\partial T^{\prime \prime}}{\partial x_j} +   u^{\prime \prime}_j  \frac{\partial \overline{T}}{\partial x_j} +
\frac{\partial u^{\prime \prime}_j T^{\prime \prime}}{\partial x_j} &= \kappa \frac{\partial^2 T^{\prime \prime}}{\partial x_j^2} 
- \frac{\partial q_{j}^{\prime \prime \, sgs}}{\partial x_j} + \frac{\partial q^{R}_j}{\partial x_j}.
\end{align}
The $1^{\rm{st}}$ and $4^{\rm{th}}$ terms on the left hand side (LHS) of the finescale momentum and temperature equations are the temporal change and  advection by the finescale, respectively. There are two additional terms on the LHS that represent interaction with the submesoscale velocity: advection by the  submesoscale ($2^{\rm{nd}}$ term) and distortion by the submesoscale gradient ($3^{\rm{rd}}$ term). On the right hand side (RHS), $\tau_{ij}^{R}$ is the residual stress and $q_j^{R}$ is the residual heat flux, same as those obtained at the submesoscale; $\tau_{ij}^{\prime\prime \, sgs}$ and $q_{j}^{\prime \prime \, sgs}$ are the finescale contributions of the model subgrid stress and heat flux, respectively.
\section{Evolution of frontal instabilities} \label{sec:frontal_instabilities}
The temporal evolution of the front under growing instabilities is summarized below before moving to the detailed analysis of the submesoscale and the finescale in the subsequent sections. 

\subsection{Symmetric instability} \label{subsec:SI}
\begin{figure*}[h!]  
\center
\includegraphics[clip=true, width=5.0in]{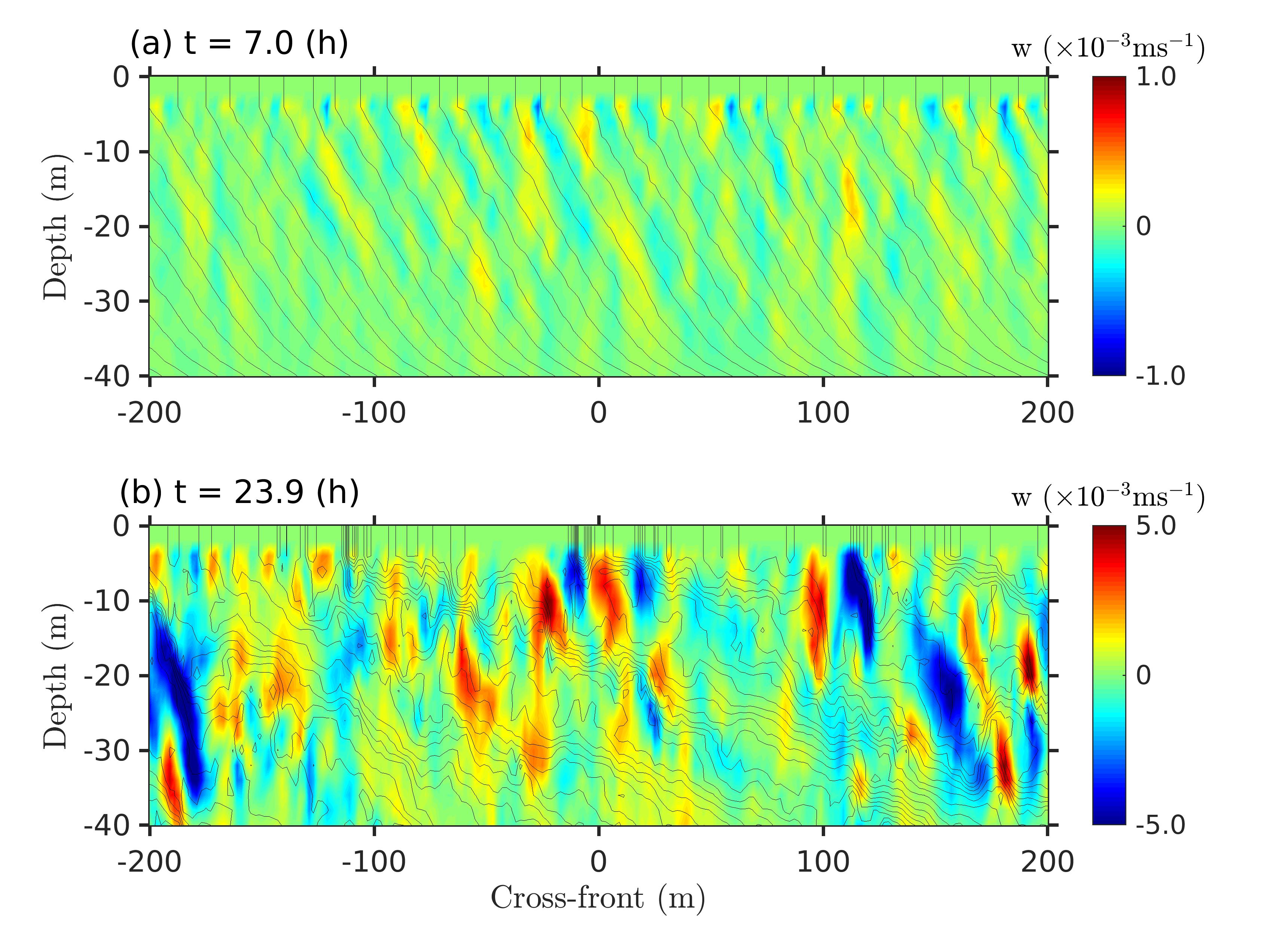}
  \caption{Evolution of SI: vertical velocity ($w$) at early (a) and late (b) stages of SI. The black solid lines represent isotherms.  
 }\label{fig:SI_evolution}
\end{figure*}

The initial potential vorticity is negative at the simulated front (Fig.~\ref{fig:initial_condition}c), making the front unstable to symmetric instability (SI) \citep{Hoskins:1974}. The evolution of SI is characterized by the formation of convection cells in the perturbation velocity which are nearly aligned with the isopycnals \citep{TaylorF:2009}. Figure~\ref{fig:SI_evolution} shows different stages of the SI evolution. Convection cells are illustrated  in Fig.~\ref{fig:SI_evolution}(a) by the bands of vertical velocity ($w$) with alternating positive and negative signs that are nearly aligned  with the isotherms (coincident with isopycnals here). When the amplitude of SI becomes sufficiently large, it undergoes secondary Kelvin-Helmholtz (KH) instability due to large vertical shear ($\partial v/ \partial z$) in the cells and KH billows form along the slanted isotherms \citep{TaylorF:2009}. Subsequently, the convection cells undergo an additional tertiary instability which appears as a lateral meandering in the $x-y$ plane with wavelength O(100) m relative to the direction in which the cells are aligned (not shown here). The tertiary instability enhances fluctuations in the cross-front vorticity component, $\omega_y^{\prime}$, and subsequently leads to the breakdown of the flow into 3-D turbulence in a process, similar to that noted by \cite{AroboneS:2015}, that commences at $ft \approx 11$  ($t = 21.8 \, \rm{h}$ in the present simulation). The turbulent stage is illustrated by the broadband, multiscale fluctuations of $w$ in Fig.~\ref{fig:SI_evolution}b. There are patches of positive and negative $w$ that are O(10) m in both  horizontal and  vertical directions. The subsequent nonlinear evolution of SI continues with the strengthening of overturns, and $w$ increasing up to $8 \, \rm{mm} \, \rm{s}^{-1}$ in magnitude.

\begin{figure*}[h!]  
\center
\includegraphics[clip=true, width=5.0in]{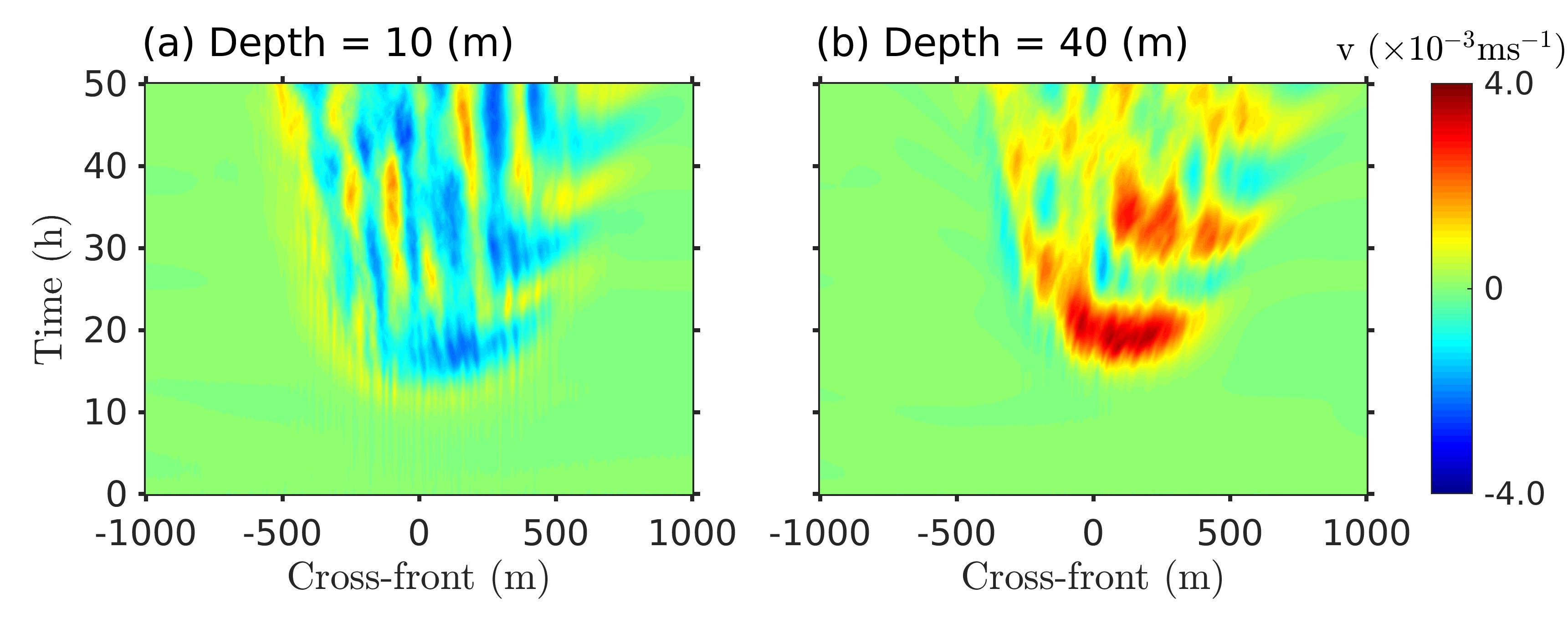}
  \caption{Ageostrophic secondary circulation develops during the evolution of SI. The time evolution of cross-front profiles of mean cross-front velocity, $\langle v \rangle_x$: (a) at 10 m depth and (b) at 40 m depth. The negative near-surface  $v$ in (a) transports water from the warmer side of the front (positive y) toward the colder side (negative y) while water at depth flows in the opposite direction in (b). 
 }\label{fig:ASC}
\end{figure*}

The growth of SI, especially the non-linear growth, leads to restratification of the front. The restratification is accompanied by the development of a secondary circulation in the y-z plane that is associated with an overall negative (light to heavy side) cross-front velocity ($v$) at the surface, oppositely-directed positive $v$ below, and associated upwelling and downwelling limbs.  Figure~\ref{fig:ASC} shows the temporal evolution of $v$, averaged in the homogeneous along-front ($x$) direction,  near the surface ($10\,\rm{m}$ depth in Fig.~\ref{fig:ASC}a) and near the bottom ($40\,\rm{m}$ depth in Fig.~\ref{fig:ASC}b) of the mixed layer. Although there is spatio-temporal variability in $v$, the behavior of $v$ with increasing time indicates that the secondary circulation has overall $v<0$ at 10 m depth that transports warm water to the colder side of the front in contrast to the oppositely-directed flow at 40 m depth with  overall $v >0$. Thus, the secondary circulation setup by SI restratifies the front, and the flow becomes stable to SI when PV becomes positive.

\subsection{Baroclinic instability} \label{subsec:BI}
\begin{figure*}[h!]  
\centering
\includegraphics[clip=true, width=5.5in]{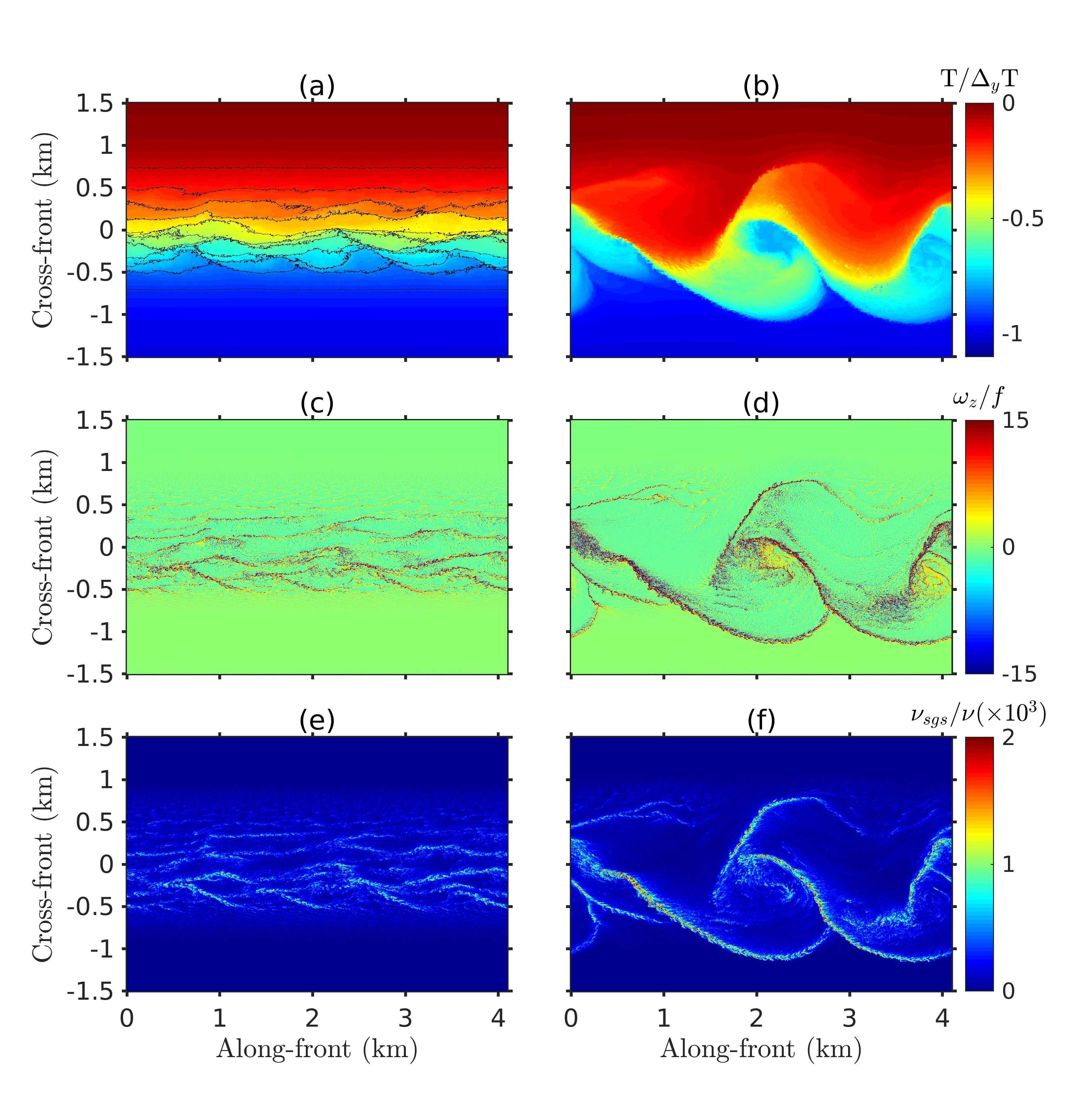}
  \caption{Frontogenesis occurs during the evolution of baroclinic instability. Non-dimensional temperature $T/\Delta_y T$, where $\Delta_y T = M_0^2L/(\alpha g)$, is plotted in panels (a) and (b); non-dimensional vertical vorticity, $\omega_z/f$, is plotted in (c) and (d); and non-dimensional subgrid viscosity, $\nu_{sgs}/\nu$, is plotted in (e) and (f). All the plots are on a horizontal ($x-y$) plane $2 \, \rm{m}$ below the surface. The figures in the left column are plotted at $t =  45 \, \rm{h}$ and those in the right column at $t = 86 \, \rm{h}$. 
 }\label{fig:frontogenesis}
\end{figure*}

As the SI subsides, baroclinic instability (BI) emerges to modulate the remnant of SI and eventually dominates the frontal instabilities. The growth of the BI mode results in the onset of large-scale meandering of the isotherms (Fig.~\ref{fig:frontogenesis}a) with  wavelength comparable to that predicted by the linear theory. Frontogenesis occurs as can be identified by the tightening of the isotherms in Fig.~\ref{fig:frontogenesis}a. Vertical vorticity plots (Figs.~\ref{fig:frontogenesis}c-d) show intensification of $\omega_z$ in the frontogenetic regions owing to locally enhanced horizontal shear. The vertical shear increases as well, following the increase in the lateral buoyancy gradient. The frontogenetic regions eventually  break down into turbulence through processes which will be quantified in later sections. The development of finescale flow component in regions undergoing frontogenesis is evident from numerous small-scale structures in Figs.~\ref{fig:frontogenesis}(b, d) and from the enhanced subgrid viscosity  in Fig.~\ref{fig:frontogenesis}f, which grows to become three orders of magnitude larger than the molecular viscosity. 

The thin regions with concentrated $\omega_z$ at the front become elongated as BI grows and transform into vortex filaments (Fig.~\ref{fig:frontogenesis}d) with lengths comparable to the wavelength of the dominant BI mode. The vortex filaments often appear in pairs -- one coming from the heavier side of the front and another from the lighter side  to join near the center line, $y = 0$.  As the vortex-filament pairs advect downstream, they roll up and organize into vorticity patches, i.e. the coherent submesoscale eddies. In Figs.~\ref{fig:frontogenesis}(b, d), two developing submesoscale eddies can be noted. The overall sense of rotation in the eddies is cyclonic. Note that the cores of the submesoscale eddies are relatively colder (i.e., heavier) than the surrounding fluid  due to entrainment of the cold water from the heavier side of the front. After forming, the submesoscale eddies continue to grow by the supply of vorticity from the filament structures. By the end of the simulation, the submesoscale eddies had grown to become $1 \, \rm{km}$ in diameter, comparable in size to the initial width of the front. 

\begin{figure*}[h!]  
\centering
\includegraphics[clip=true, width=6.0in]{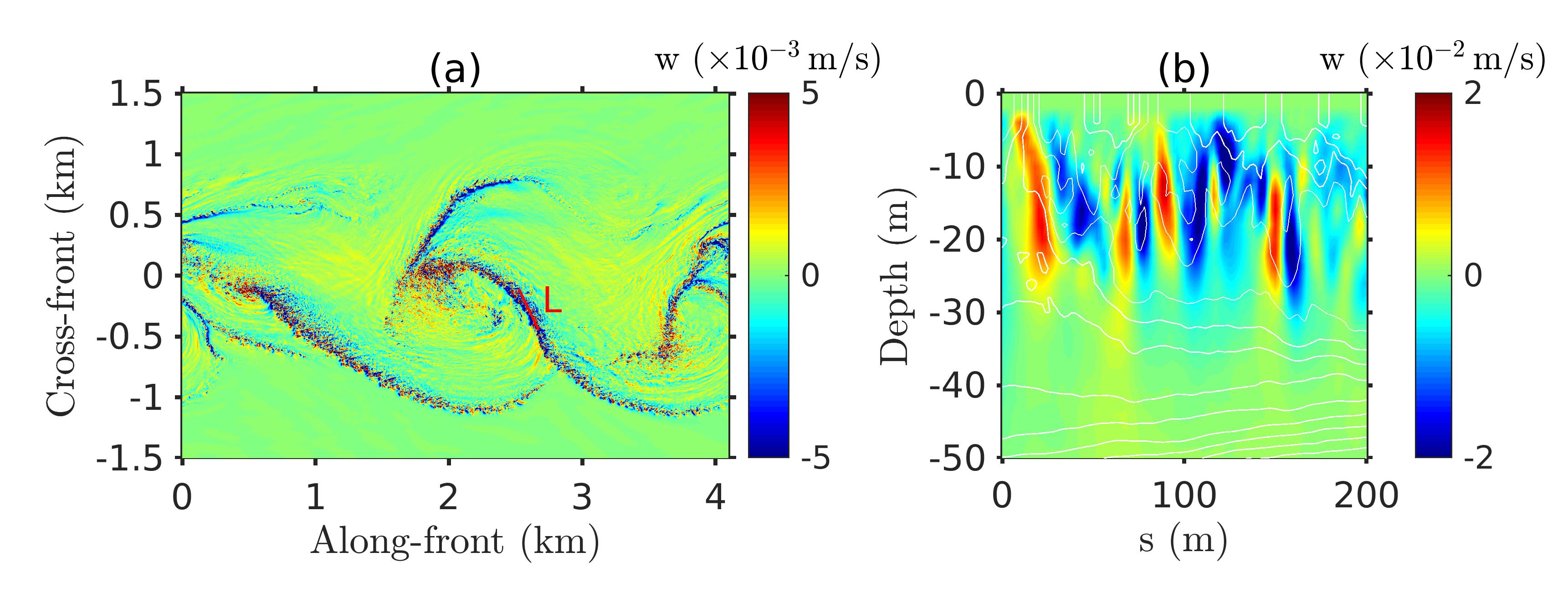}
  \caption{ (a) Vertical velocity on a horizontal plane at $10 \, \rm{m}$ depth at $t = 86 \, \rm{h}$. (b) Vertical velocity on a vertical cross-section along the red solid line shown in the panel (a); $s$ is the distance measured along the line, moving in the positive y-direction. The solid white lines in  panel (b) represent isotherms.
 }\label{fig:vertical_velocity}
\end{figure*}

Vertical vorticity is concentrated in the coherent structures, i.e. the vortex filaments and the coherent submesoscale eddies. The value of local Rossby number ($\hat{Ro} = |\omega_z|/f$) in the coherent structures is as large as 50 \hl{(Fig.~\ref{fig:frontogenesis}d shows $\omega_z/f \in [-15, 15]$ for better visualization)} and suggests significant ageostrophy and loss of balance in these structures. Moreover, the concentration of large subgrid viscosity (Fig.~\ref{fig:frontogenesis}f) indicates strong turbulence in the coherent structures. The ageostrophic regions develop large vertical velocity. In Fig.~\ref{fig:vertical_velocity}a, $w$  is plotted on a horizontal plane at $10 \, \rm{m}$ depth; its similarity with the $\omega_z$-field (Fig.~\ref{fig:frontogenesis}d) is apparent with large-$w$  regions concentrated within the coherent structures. The magnitude of $w$  becomes  as large as $0.02 \, \rm{m}\,\rm{s}^{-1}$ or about $0.35U_0$. Figure \ref{fig:vertical_velocity}b shows $w$ on a cross-section whose vertical coordinate is the depth and horizontal coordinate is length ($s$) along a vortex filament (the  line $L$ in Fig.~\ref{fig:vertical_velocity}a). The $w$-field (Fig.~\ref{fig:vertical_velocity}b) within the vortex filament reveals bands of positive and negative $w$ with 20 m characteristic length scale. 

The BI is effective in restratifying the front and the process is facilitated by an ASC \citep{Hoskins:1982}. Although large patches with coherent $w$ (especially upwelling) can be identified in Fig.~\ref{fig:vertical_velocity}a, the finescale of the eddies  obscures the ASC. The probability density function (PDF)  of $w$  shows asymmetry for magnitudes greater than $2 \, \rm{mm\,s^{-1}}$ in favor of downwelling motion (not shown here).  However, the corresponding probability of these large-$w$ regions is quite small and the presence of net downwelling motion in the coherent structures is not obvious. Indeed, separation of the finescale from the submesoscale is necessary to clarify vertical transport as is done in section~\ref{sec:scale_separation}.

\begin{figure*}[h!]
 \centering
 \includegraphics[clip=true, width=5.0in]{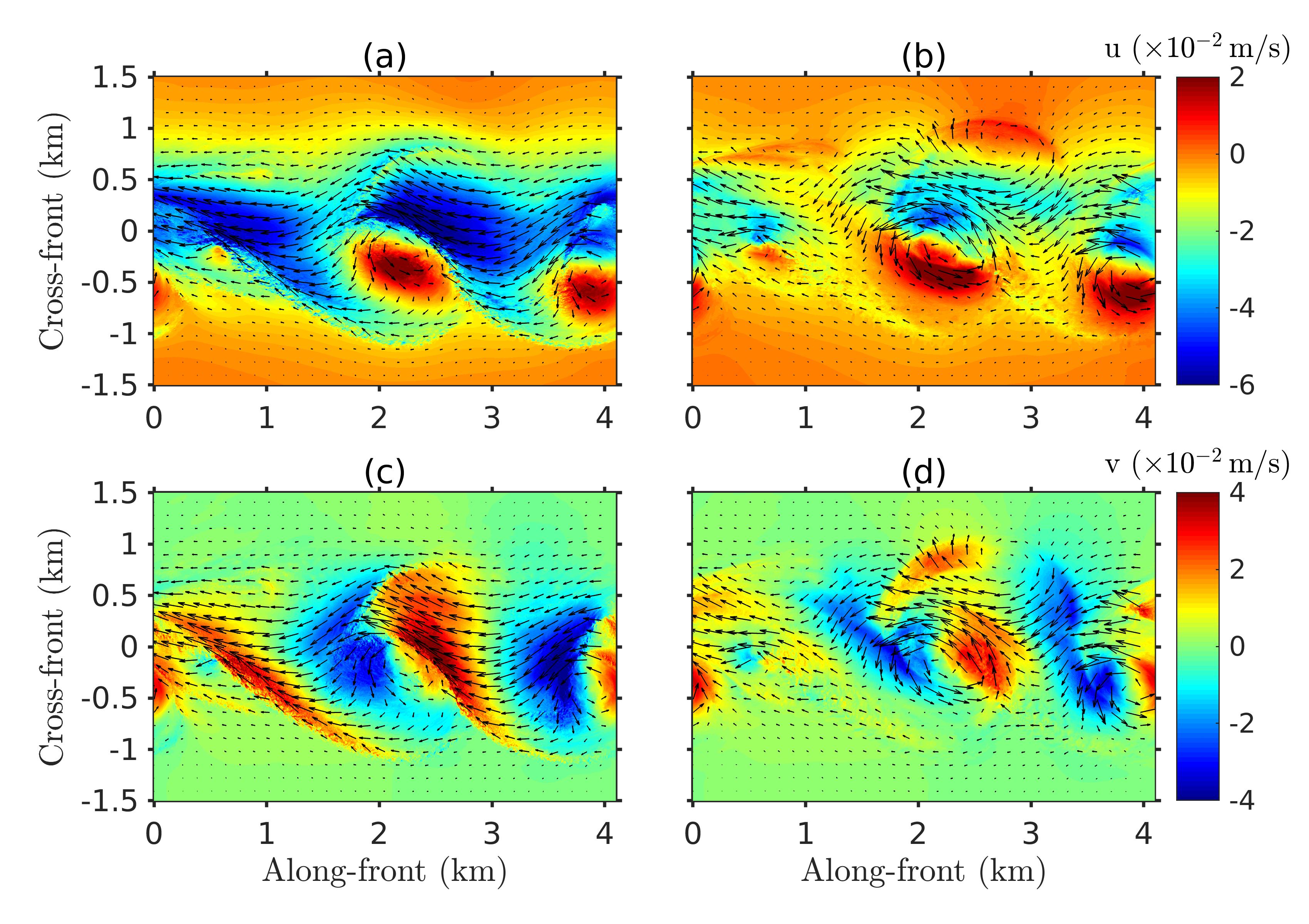}
 \caption{\hl{Deformation of the frontal jet due to the coherent filaments and eddies depicted at $t = 86 \, \rm{h}$}. Along-front velocity component, $u$, at 10 m and 30 m depths are plotted in panels (a) and (b), respectively; similarly, cross-front velocity component, $v$, at 10 m and 30 m depths are plotted in panels (c) and (d). The arrows show the horizontal velocity vectors.}
 \label{fig:frontal_jet}
\end{figure*}

The growth of BI modifies the frontal jet as can be seen from  the velocity contours and vectors of Fig.~\ref{fig:frontal_jet}. The width of the jet near the surface increases, and it becomes more energized. The initial range of $u$ from \hl{$-5\times 10^{-2}$ to 0  $\rm{m\,s^{-1}}$} expands with increasing time and notably includes positive $u$ up to \hl{$2 \times 10^{-2} \, \rm{m\,s^{-1}}$}. The cross-front velocity ($v$), initially zero, becomes as large as $u$. Clearly, a considerable amount of kinetic energy is transferred to the swirling motion of the submesoscale eddies. These eddies speed up the jet on the lighter side of the front and slow it down on the heavier side. The eddies span the entire mixed layer depth and the modification of the jet below mid-depth is relatively strong since the initial $u$ is small relative to the surface. In this paper, we use the term `front' to denote the region where there is significant horizontal velocity. The largest density gradients are located mainly in the coherent structures and found to be confined within the front.

\subsection{A characteristic lateral dimension for the spatially coherent finescale}  \label{subsec:filterscale}
\begin{figure}[h!]  
\centering
\includegraphics[clip=true, width=4.0in]{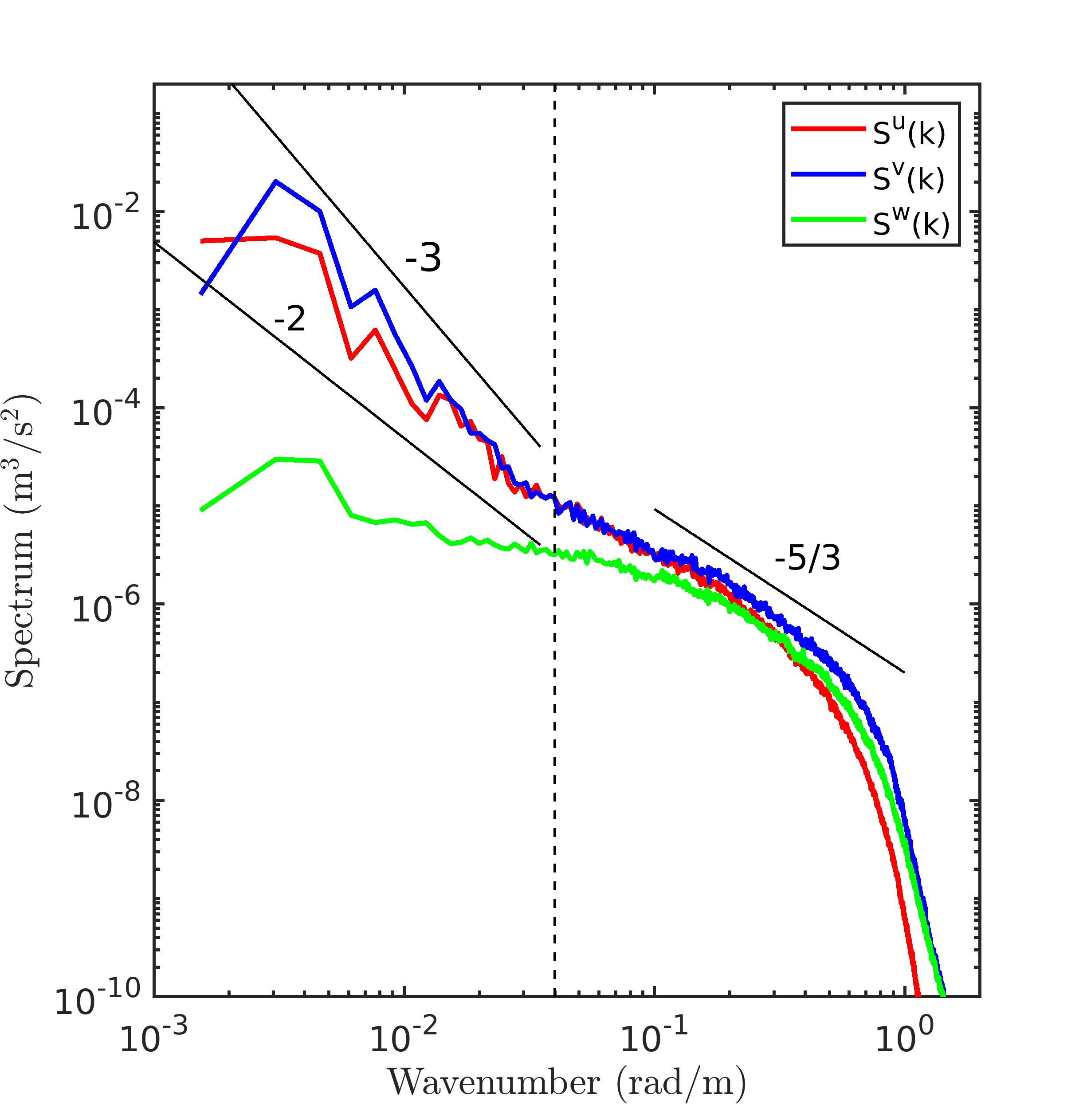}
  \caption{Power spectra of along-front velocity ($S^u(k)$), cross-front velocity ($S^v(k)$), and vertical velocity ($S^w(k)$) at $20 \, \rm{m}$ depth and $t = 86 \,\rm{h}$. 
  The dashed vertical line is plotted at $k = 0.04 \, \rm{rad\,m^{-1}}$.
 }\label{fig:energy_spectrum}
\end{figure}

Flow visualization  shows formation of thin filament structures due to frontogenesis as BI evolves. Interrogation of the velocity and temperature fields in physical space reveals  a characteristic lateral dimension of $O(100)$ m for the spatially-coherent finescale that is associated with  the vortex filaments. The $O(100)$ m scale of  frontal arrest   also has an imprint on the spectra. The power spectra of the along-front velocity $u$, cross-front velocity $v$ and vertical velocity $w$ at 20 m depth are shown in Fig.~\ref{fig:energy_spectrum}. The spectra are computed as follows:
\begin{align}
 S^u(k) &= \frac{1}{2} \left<|\hat{u}(k)|^2 \right>_y, \quad S^v(k) = \frac{1}{2} \left<|\hat{v}(k)|^2 \right>_y, \nonumber \\
 S^w(k) &= \frac{1}{2} \left<|\hat{w}(k)|^2 \right>_y, 
\end{align}
where $k$ is the wavenumber in the along-front direction; the caret denotes the Fourier transform and $\left< \cdot \right>_y$ denotes averaging in the lateral ($y$)  direction over all computational grid  lines between $-1.2 < y < 1.2 \, \rm{km}$. 

The spectra (Fig.~\ref{fig:energy_spectrum}) show  qualitative changes with increasing $k$. At low $k$, the kinetic energy in the vertical motion is much smaller than the energy in the horizontal motions, indicating a predominantly two-dimensional flow at these scales whereas,  at higher  $k$, the energy in the vertical motion becomes comparable to those in the horizontal motions, indicating  three-dimensional turbulence. The spectra ($S_u$ and $S_v$) of the horizontal velocity components also change  from a steeper slope   to a slope closer to $k^{-5/3}$ at $k$ corresponding to $O(100)$ m. 

A choice needs to be made for $k_c$, the cutoff wavenumber of the filter that determines the split between submesoscale and finescale. From the preceding discussion, $k_c$ corresponds to $O(100)$ m. Table~\ref{tab:energy_content} shows the kinetic energy content $E^>$  in wavenumber $k > k_c$ and the distribution of $E^>$ among the three velocity components for different choices of $k_c$. The table shows an increase in the relative contribution of the vertical motion ($E^>_w$) with increasing $k_c$. At the value  $k_c = 0.04 \, \rm{rad\,m^{-1}}$ chosen for the decomposition, $E^>_w$ becomes more than 20\% and further increase of $k_c$ leads to relatively gradual change in $E^>_w$. {\em A posteriori}, we find that our choice of $k_c$ leads to a good separation of the submesoscale dynamics from the turbulent finescale.

A computational restriction is worth noting.  It is clear that the coherent structures also introduce an important length scale to the problem -- the width of the filament -- which is much smaller, $O(100)$ m, compared to the wavelength of the baroclinic instability mode.  This length scale restricts the horizontal grid size of submesoscale eddy-resolving simulations to be at least $O(100)$ m and of turbulence-resolving simulations to $O(10)$ m or less.
\section{Submesoscale} \label{sec:scale_separation}
The classical decomposition of the flow between an along-front mean and a fluctuation is not adequate to understand the  role of the coherent submesoscale and its interaction with 3-D turbulence. The mean flow computed by averaging the flow in the along-front direction would overlook the spatio-temporal coherence of the submesoscale, as well as its imprint on the finescale. Here, the finescale is separated from the large-scale flow by using a low-pass filter: a two-dimensional Lanczos filter (Eq.~\ref{eq:2DLanczos}) with $a = 2$ and $k_c = 0.04 \, \rm{rad\,m^{-1}}$. The Lanczos filter successfully extracts the coherent structures (as will be shown) while providing a sharp cutoff in the wavenumber space; structures with $k > k_c$ are removed while the energy content of the larger structures remains largely unaffected. The value of \hl{the cutoff wavenumber}, $k_c$, is  motivated by the change in flow anisotropy (quasi-2D to 3D motions) and spectral slopes observed at $O(100)$ m that was discussed in section  \ref{subsec:filterscale}. The overall performance of the filter in separating the scales is assessed by the visualization of Fig.~\ref{fig:scale_separation} where submesoscale fields, denoted by $\overline{(\cdot)}$,  and finescale components, denoted by ${(\cdot)}^{\prime\prime}$, are shown on a horizontal plane. Qualitatively, the filter provides a good representation of the finescale in the flow while preserving the large-scale flow features. It is clear that there is considerable energy corresponding to the swirling flow that is spread across the submesoscale while  the finescale is localized.

\subsection{Vertical transport}
The front develops large \hl{vertical velocities} as was shown by Fig.~\ref{fig:vertical_velocity} and accompanying discussion. However, the $w$-field does not show distinct regions where upwelling and downwelling motions occur. The underlying structure of the vertical transport can be better understood using the submesoscale component of $w$.   

In Fig.~\ref{fig:scale_separation}c, the submesoscale vertical velocity, $\overline{w}$, is plotted at $10 \, \rm{m}$ depth. The $\overline{w}$-field reveals large patches with predominantly positive and negative values, {\em separating} regions with upwelling and downwelling motions. The downwelling motion, as will be shown subsequently, transports high density fluid to the interior. The organization of the patches with $\overline{w} < 0$ and $\overline{w} > 0$ is connected to the coherent structures. Negative vertical velocity is concentrated along the filaments and the magnitude is relatively stronger in regions where filaments curve into the coherent eddies. The solid black contours of $\overline{w} = -0.1 \, \rm{mm\,s^{-1}}$ estimate the edges of the $\overline{w} < 0$ regions. On the other hand, the upwelling regions, $\overline{w} > 0$, are spatially extensive and occupy a much larger frontal area relative to $\overline{w} < 0$ regions. In Fig~\ref{fig:scale_separation}c, $\overline{w} > 0$ regions are identified by the black dash-dot lines corresponding to  $\overline{w} = 0.1\,\rm{mm\,s^{-1}}$. Thus, at the submesoscale, the regions with positive and negative vertical velocities become separated, exposing the underlying structure of the upwelling and downwelling motions.   

Similarly, at the submesoscale, regions of positive and negative vertical vorticity are well separated (Fig.~\ref{fig:scale_separation}g). The figure shows that $\overline{\omega}_z$ is positive (cyclonic) in the filament structures and in the coherent eddies, and has magnitude of 5-7$f$; negative $\overline{\omega}_z$ reaches only $1.2f$ in magnitude. The finescale vorticity (${\omega}_z^{\prime\prime}$) has large magnitude up to $50f$ which, unlike the submesoscale, is comparable for both negative and positive components. The intertwined positive and negative ${\omega}_z^{\prime\prime}$ values cancel out in the finescale, and it is the submesoscale vorticity that dictates the rotation of the turbulent filament structure as it rolls-up anticlockwise to form a cyclonic eddy. 

We now take a deeper look into downwelling and upwelling motions through the following decomposition of $w$:
\[
 w = w^{+} + w^{-};
 \begin{cases}
  w^{+} = 0, w^{-} \neq 0, w < 0 \\
  w^{+} \neq 0, w^{-} = 0, w \geq 0.
 \end{cases}
\]
Thus, $w^{-}$ represents the negative part of $w$, while $w^{+}$ represents the positive part including $w=0$. 

\begin{figure*}[h!]
 \centering
 \includegraphics[clip=true, width=6.0in]{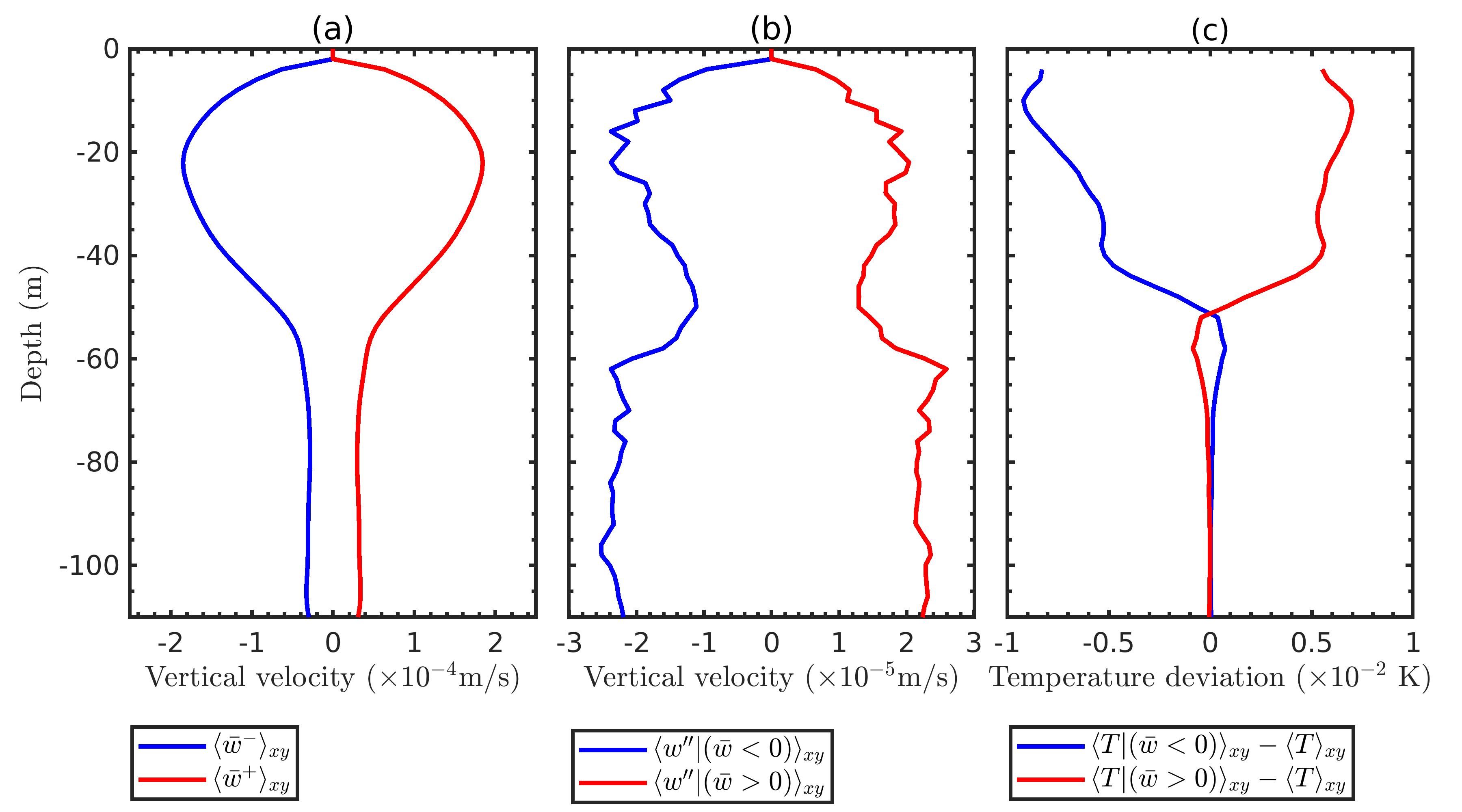}
 \caption{Vertical transport at the front at $t = 86 \, \rm{h}$: (a) frontal averages of positive and negative parts of the vertical velocity, $\langle \bar{w}^{+} \rangle_{xy}$ and $\langle \bar{w}^{-} \rangle_{xy}$; (b) frontal averages of the finescale vertical velocity sampled in regions with $\overline{w} > 0$ and $\overline{w} < 0$; (c) frontal averages of $T$-deviation sampled in regions with $\overline{w} > 0$ and $\overline{w} < 0$,  where the deviation is measured with respect to the overall horizontal average, $\langle T \rangle_{xy}$.}
 \label{fig:vertical_profiles}
\end{figure*}

The frontal averages of the vertical velocity and the temperature are plotted in Fig.~\ref{fig:vertical_profiles}. The average has been calculated over the region most influenced by BI: the entire along-front length ($L_x$) and  $-1.2 \leq y \leq 1.2\,\rm{km}$ across the front. The vertical profiles of  Fig. \ref{fig:vertical_profiles}(a) show that  $\langle \overline{w}^{+} \rangle_{xy}$ and $\langle \overline{w}^{-} \rangle_{xy}$ are nearly equal in magnitude, but opposite in sign (the difference in magnitudes is at least one-two orders smaller than either) in conformity with the continuity requirement at the submesoscale. Both $\langle \overline{w}^{+} \rangle_{xy}$ and $\langle \overline{w}^{-} \rangle_{xy}$, which are zero near the surface due to the no-penetration boundary condition, increase to their maximum values at the depth of $\approx 20\,\rm{m}$ and then decrease with increasing depth. However, they do not become zero at the mixed layer depth $H = 50\,\rm{m}$. The non-zero $w$ in the thermocline is due to internal waves and found to have \hl{a different spatial structure} than that in the mixed layer.

The overall upwelling and downwelling  transport in the mixed layer is dominated by the submesoscale component, $\overline{w}$.  Figure~\ref{fig:vertical_profiles}b shows averages of the finescale, conditioned on the sign of the submesoscale $w$. In the region with $\overline{w} < 0$, the net effect of the finescale is downwelling (average of $w^{\prime \prime}$ in that region is negative) and, in the region with $\overline{w} > 0$, the net effect is upwelling. The vertical upwelling and downwelling due  to $w^{\prime \prime}$ of the finescale (Fig.~\ref{fig:vertical_profiles}b) is found to be much smaller, at least 7-8 times smaller than those by the submesoscale velocity. Therefore, measurement of the submesoscale vertical velocity is sufficient to obtain the net upwelling/downwelling.

The instability correlates temperature (density) with the vertical velocity. Figure~\ref{fig:vertical_profiles}c shows profiles of averaged $T$, conditioned on the sign of $\overline{w}$. In the regions with positive (negative) $\overline{w}$,  the conditional mean of temperature is higher (lower) than the total average. Thus, the submesoscale motions transport high-density (low-$T$) fluid near the surface to the interior of the mixed layer and, consequently, the front restratifies. The correlation (not plotted) between the density and the vertical velocity is negative indicating the conversion of potential energy to kinetic energy. 

Comparison of the $\overline{\omega}_z$ and $\overline{w}$ fields in Fig.~\ref{fig:scale_separation} shows a correspondence between the regions with strong positive $\overline{\omega}_z$ and negative $\overline{w}$. It is tempting to use $\overline{\omega}_z > 0$ as a predictor of downwelling. However, the computed correlation between $\overline{\omega}_z$ and $\overline{w}$ is small. Also, the vertical downwelling ($w^-$) computed by conditioning on positive $\overline{\omega}_z$ is found to be substantially smaller than the net  $w^-$ at the front. The reason is that the eddy core with positive $\overline{\omega}_z$ has a substantial area of upwelling, not downwelling.

The downwelling and upwelling motions at the front can be considered as parts of an ageostrophic secondary circulation (ASC). In simplified models of fronts where the along-front variation is averaged or neglected, the ASC is a restratifying 2-D ($y-z$) circulation that is anticlockwise looking in the direction of the jet. The ASC transports near-surface  water from the light ($y>0$) to the heavy ($y< 0$) side while dense water subducts from the heavy side and moves toward the light side at depth.  

The ASC which develops during the evolution of BI is in fact 3-D owing to the along-front variability associated with the coherent submesoscale. The along-front variability of the $y-z$ circulation is depicted in Fig.~\ref{fig:BI_SAC} by contrasting the circulation among three $y-z$ cuts (whose intersections with  $z = 10$ are marked as S1, S2 and S3 in Fig. \ref{fig:scale_separation}) of a submesoscale eddy. If we define {\em back-to-front} of the eddy to be in the direction (negative $x$) of the average jet velocity, S3 is at the back, S2 is near the center, and S1 is at the front of the eddy. Note that, consistent with the anticlockwise $x-y$ circulation of the submesoscale eddy, surface water is transported in the negative $y$-direction at S1 (Fig.~\ref{fig:BI_SAC}a) and the positive $y$-direction at S3 (Fig.~\ref{fig:BI_SAC}c). The $y-z$ circulation at S1 (Fig.~\ref{fig:BI_SAC}a) shows that there is downwelling in the filament ($-700 < y < -600$ m) and predominantly upwelling in the eddying region ($-300 < y < 500$ m). On the other hand, the circulation at S3 (Fig.~\ref{fig:BI_SAC}c) shows substantial downward motion. S2 (Fig.~\ref{fig:BI_SAC}b) is a $y-z$ section through the eddy core.  Within the eddy, the temperature contours at the bottom of the mixed layer and in the vicinity of $y=0$ indicate isopycnal doming under the influence of low pressure, and the fluid has radially-outward horizontal velocity. At S2, there is both upwelling and downwelling with the former in the eddy core and the latter at the back of of the eddy. To summarize, the organization of the secondary circulation in the $y-z$ plane is associated with the coherent structures at the front, with upwelling dominant in the forward and central  regions of the submesoscale eddy and downwelling in the aft-regions and the filaments.  
 
\subsection{3D structure of the coherent submesoscale}
The downwelling of high-density water and the upwelling of low-density water are three-dimensional processes, each dominating in  different parts of the front and mediated by the coherent structures. The three-dimensionality of the   coherent structure helps shed light  on the vertical exchange.

\begin{figure*}[h!]
   \centering
   \includegraphics[clip=true, width = 6.0in]{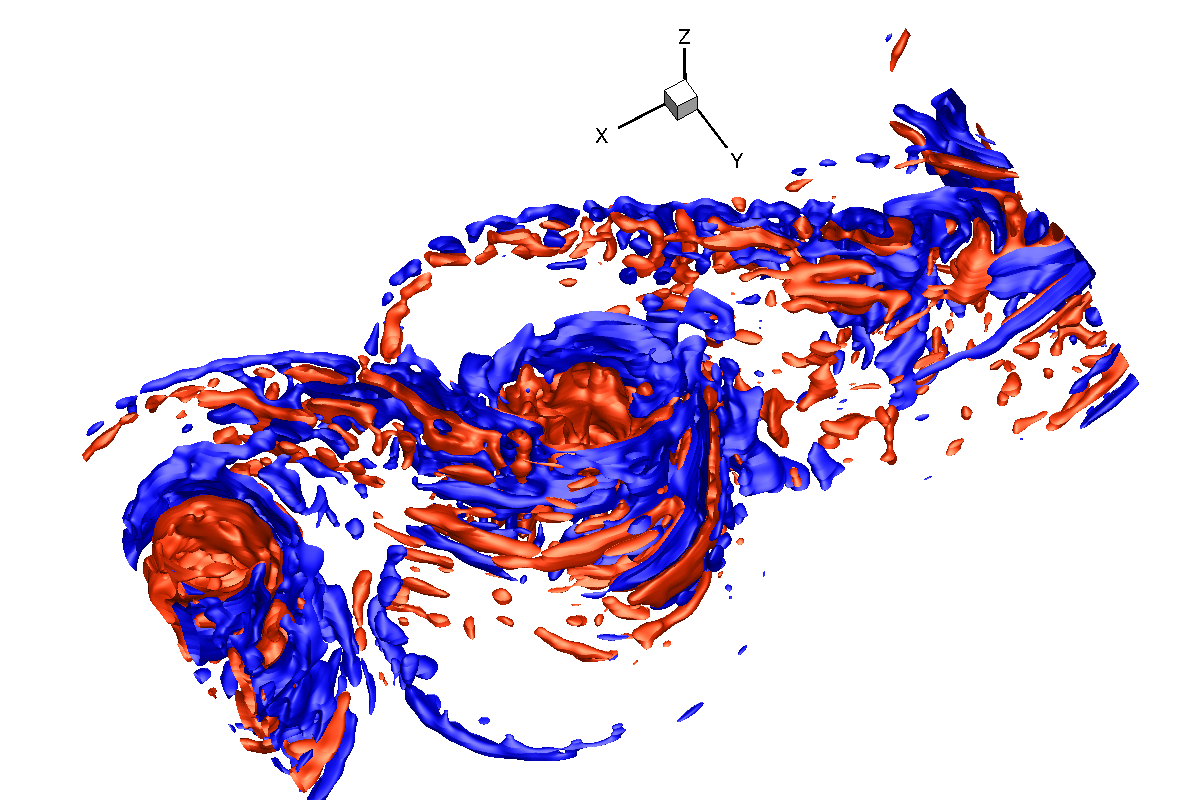} 
   \caption{Three-dimensional visualization of the coherent structures using the $Q$-criterion at $t = 86 \, \rm{h}$. The iso-surfaces are plotted at $Q/f^2 = $ 0.4 (red) and -0.4 (blue).}  
   \label{fig:Q_LS_n13200}
\end{figure*}

Here, the coherent structure is extracted using the $Q$ criterion defined as
\begin{equation}
 Q = \frac{1}{2} \left( \overline{\Omega}_{ij}^2 - \overline{S}_{ij}^2 \right); \quad \overline{\Omega}_{ij} = \frac{1}{2}\left( \frac{\partial \overline{u}_i}{\partial x_j}-\frac{\partial \overline{u}_j}{\partial x_i} \right), \; \overline{S}_{ij} = \frac{1}{2}\left( \frac{\partial \overline{u}_i}{\partial x_j}+\frac{\partial \overline{u}_j}{\partial x_i} \right),
\end{equation}
where $\overline{\Omega}_{ij}$ and $\overline{S}_{ij}$ are the rotation and strain tensors of the submesoscale, respectively.  $Q >0 $ signifies rotation-dominated flow  while $Q < 0 $ signifies the domination of strain. Iso-surfaces of $Q$ (Fig.~\ref{fig:Q_LS_n13200})  show that the coherent structures consist of both rotation- and strain-dominated regions which are  organized in layers around the coherent submesoscale eddies. It can also be noted that the filament structure at the heavier side ($y<0$) of the front is shallow at its origin, and its depth of influence increases as one moves along this structure towards the eddy; very close to the eddy and in the region surrounding the eddy, the filaments influence the whole of the mixed layer. The filamentary structures at the lighter side ($y>0$)  of the front seem to affect the entire mixed layer depth. 

Thus, the separation of scales of motion into the submesoscale and the finescale provides a better understanding of the vertical velocity organization and the vertical transport at the baroclinic frontal instability. The submesoscale vertical velocity field shows well-separated regions where either negative or positive $w$ dominates. Moreover, the vertical velocity field is connected to the coherent structures. Strong downwelling motions develop at vortex filaments and in some portions of the submesoscale eddies that adjoin the filaments.
\section{Kinetic energy of the submesoscale and the finescale} \label{sec:kinetic_energy}

In this section, we  assess the relative contributions of the submesoscale and the finescale to the kinetic energy, quantify the dominant balances in the transport equations, Eq.~\ref{eq:sms_eqn} and Eq.~\ref{eq:fs_eqn} for the submesoscale and finescale, respectively, and make explicit the interaction between turbulence -- the finescale -- and the submesoscale.

\begin{figure}[h!]  
\centering
\includegraphics[clip=true, width=4.0in]{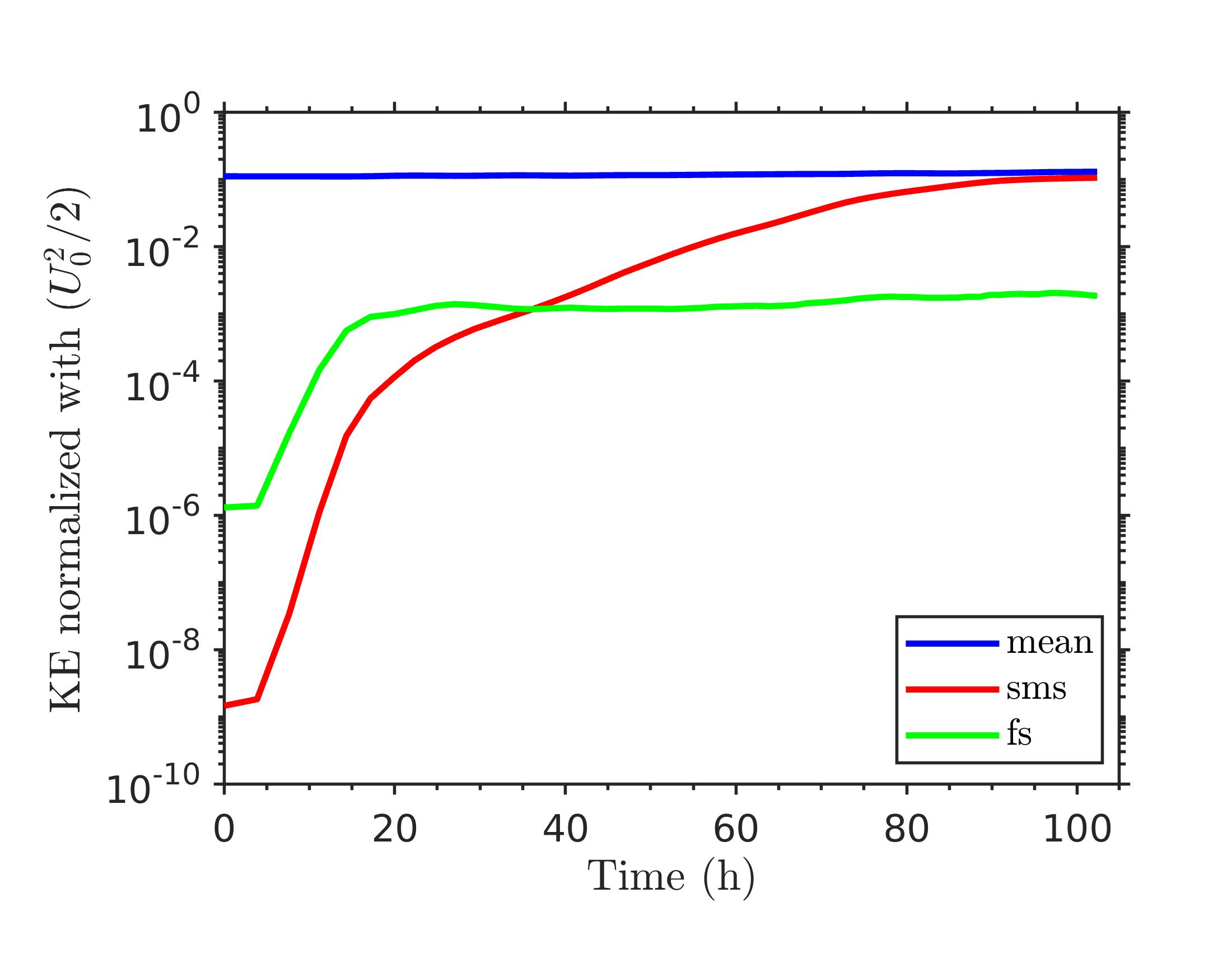}
  \caption{Plot of the bulk values of mean kinetic energy, $\langle u_i \rangle_x^2/2$, submesoscale fluctuation kinetic energy, $\left(\overline{u}_i-\langle \overline{u}_i \rangle_x \right)^2/2$, and finescale fluctuation kinetic energy, $ \left( u_i^{\prime \prime}-\langle u_i^{\prime \prime} \rangle_x \right)^2/2$. The bulk values are obtained by volume averaging over the horizontal extent of the front and  the entire mixed layer depth.    
 }\label{fig:ke_time}
\end{figure}

The submesoscale velocity, as defined here, includes the along-front ($x$) average which at $t=0$ is the velocity of the initially-balanced geostrophic jet. In order to focus on the fluctuations, we subtract the along-front average before computing the KE and contrast the evolution of the overall (volume-averaged) fluctuation energy in the submesoscale and the finescale components in  Fig.~\ref{fig:ke_time}. Initially, the mean KE dominates and the KE of the small-amplitude broadband fluctuations introduced at $t =0$ resides mainly in the finescale. Figure~\ref{fig:ke_time} shows that both submesoscale and finescale KE increase during the growth of SI. When the instability becomes nonlinear at $t \approx 15$ h, the finescale KE saturates, but the submesoscale KE  keeps increasing. At $t \approx 40$ h, the submesoscale KE overtakes the finescale KE. The subsequent evolution is dominated by BI which increases the 
submesoscale KE until it finally saturates at a value comparable to the mean KE. Although the volume-averaged KE of the finescale in the frontal region does not change significantly during the evolution of BI, there is substantial spatio-temporal variability as the  finescale, concentrated in the coherent filament structures and eddies, interacts with the submesoscale. How the submesoscale and finescale flows evolve and interact is investigated here by studying the KE budgets at the two scales. An outstanding question is what is {\em the source} of the finescale KE that counteracts its dissipation?

The submesoscale-KE equation, derived by multiplying Eq.~\ref{eq:sms_eqn} with $\overline{u}_i$, is given by 
\begin{align} \label{eq:les_mean_ke}
 \frac{\partial}{\partial t} \frac{\overline{u}_i \overline{u}_i}{2} =  -\frac{\partial \overline{T_j}}{\partial x_j} + \overline{B} - \overline{\mathcal{E}} -\overline{\mathcal{E}}^{sgs} - \overline{\mathcal{E}}^{R},
\end{align}
where $\overline{T_j}$ is the transport term,
\begin{align}
 \overline{T_j} =  \frac{\overline{u}_j\overline{u}_i\overline{u}_i}{2} + \frac{\overline{u}_j \overline{p}}{\rho_0} - \nu \frac{\partial }{\partial x_j} \frac{\overline{u}_i \overline{u}_i}{2}+\overline{u}_i(\overline{\tau}_{ij}^{sgs} + \tau_{ij}^{R}),
\end{align}
$\overline{B} = \alpha \overline{T}\,\overline{w} g$ is the submesoscale buoyancy production term, $\overline{\mathcal{E}} = \nu (\partial \overline{u}_i/\partial x_j)^2$ is the molecular dissipation, and  $\overline{\mathcal{E}}^{sgs} = -\overline{\tau}_{ij}^{sgs} (\partial \overline{u}_i/\partial x_j)$ is the dissipation corresponding to the subgrid stresses. The term $\overline{\mathcal{E}}^{R} = - {\tau}_{ij}^{R} (\partial \overline{u}_i/\partial x_j)$ that arises from the effect of the submesoscale velocity gradient on the residual stresses will be shown later to act as a loss of submesoscale KE. Note that the residual stress, $\tau_{ij}^{R} = \overline{u_i u_j} - \overline{u}_i \overline{u}_j$, which arises from the submesoscale-finescale decomposition is explicitly computed after applying the Lanczos filter and is different from the subgrid stress, $\tau_{ij}^{sgs}$, which is modeled in the LES approach. 

Similar to the KE budget at the submesoscale, the KE budget at the resolved finescale can be derived by multiplying Eq.~\ref{eq:fs_eqn} with $u^{\prime \prime}_i$. The KE transport equation at the finescale is given by
\begin{align} \label{eq:les_turb_ke}
 \frac{\partial}{\partial t}\frac{u^{\prime \prime}_i u^{\prime \prime}_i}{2} = -\frac{\partial T^{\prime \prime}_j}{\partial x_j} + Tr + B^{\prime \prime} - \mathcal{E}^{\prime \prime} - \mathcal{E}^{\prime \prime \, sgs} - \mathcal{E}^{\prime \prime \, R}, 
\end{align}
where $T^{\prime \prime}_j$ is the transport term,
\begin{align}
 T^{\prime \prime}_j = \frac{\bar{u}_j u^{\prime \prime}_i u^{\prime \prime}_i}{2} + \frac{u^{\prime \prime}_j u^{\prime \prime}_i u^{\prime \prime}_i}{2} + \frac{p^{\prime \prime}u^{\prime \prime}_j}{\rho_0} - \nu \frac{\partial}{\partial x_j} \frac{u^{\prime \prime}_iu^{\prime \prime}_i}{2} + u^{\prime \prime}_i (\tau_{ij}^{\prime \prime \, sgs} - \tau_{ij}^{R}),
\end{align}
$Tr = -u^{\prime \prime}_i u^{\prime \prime}_j (\partial \bar{u}_i/\partial x_j)$ is the {\em transfer term} that  represents  the action of submesoscale velocity gradients (including shear)  on the finescale, $B^{\prime \prime} = \alpha T^{\prime \prime} w^{\prime \prime} g$ is the finescale buoyancy production, and $\mathcal{E}^{\prime \prime} = \nu (\partial u^{\prime \prime}_i/\partial x_j)^2$ and $\mathcal{E}^{\prime \prime \, sgs} = - \tau_{ij}^{\prime \prime \, sgs} (\partial u^{\prime \prime}_i/\partial x_j)$ are the dissipation of finescale KE due to molecular viscosity and subgrid stresses, respectively. $\mathcal{E}^{\prime \prime \, R} = \tau_{ij}^{R} (\partial u^{\prime \prime}_i/\partial x_j)$ term arises from the interaction of the residual stresses with the finescale velocity gradient; it will be shown later to act as a production term for the finescale KE. 

\begin{figure}[h!]
   \centering
   \includegraphics[clip=true, width=6.0in]{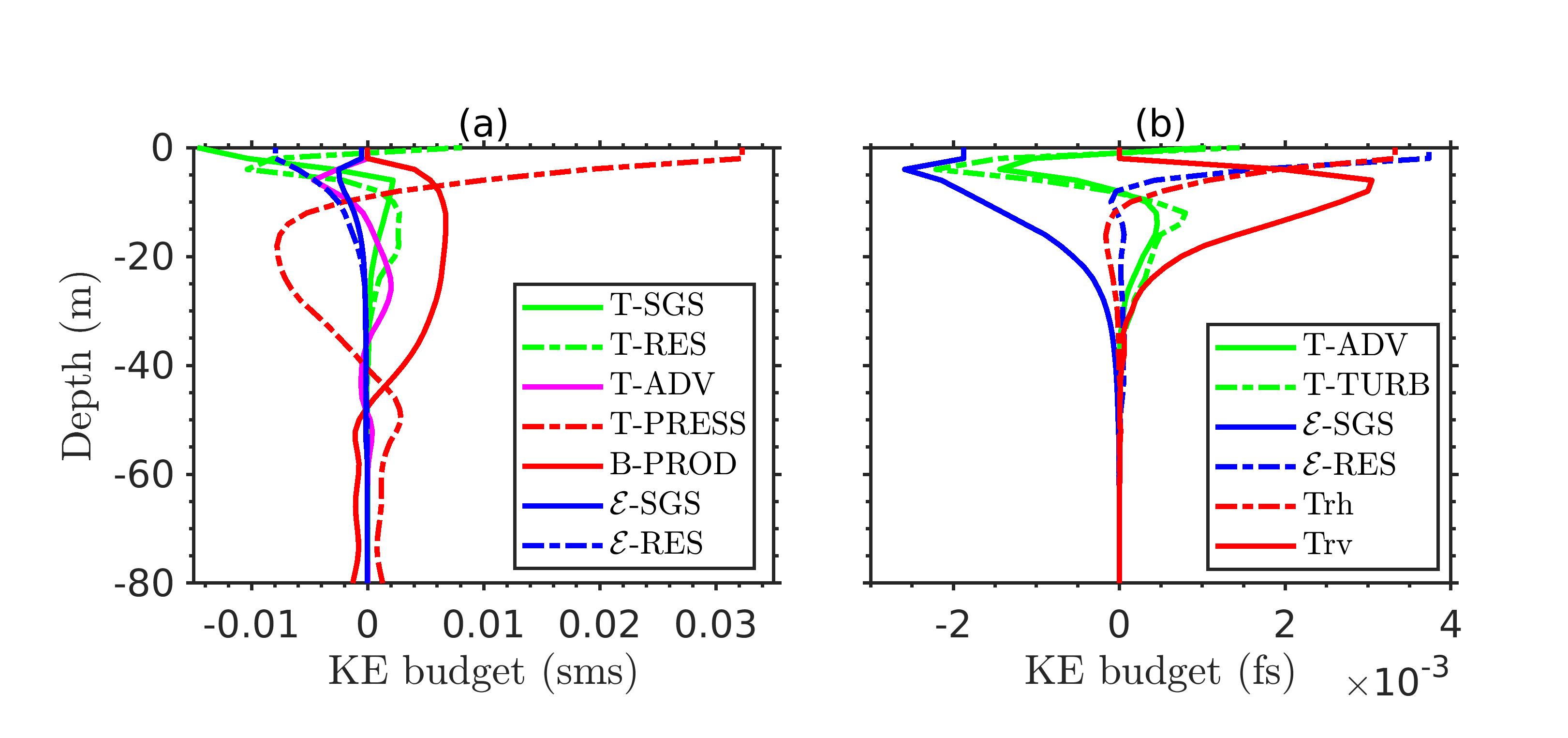} 
   \caption{Submesoscale (a) and finescale (b) kinetic energy (KE) budgets at $t = 79.7 \, \rm{h}$. Each term is a horizontal frontal average and is normalized by $U_0^2 f$. At submesoscale, the plotted terms are 
   advective transport, T-ADV = $-(1/2)\partial{(\overline{u}_j \overline{u}_i \overline{u}_i)}/{\partial x_j}$, 
   pressure transport, T-PRESS = $-(1/\rho_0)\partial(\overline{u}_j\overline{p})/\partial x_j$, 
   subgrid-stress transport, T-SGS = $-\partial(\overline{u}_i \overline{\tau}_{ij}^{sgs})/\partial x_j$, 
   residual-stress transport, T-RES = $-\partial(\overline{u}_i \overline{\tau}_{ij}^{R})/\partial x_j$, 
   buoyancy production, B-PROD = $\overline{B}$ = $\alpha \overline{T} \overline{w} g$, 
   subgrid dissipation, $\mathcal{E}$-SGS = $-\overline{\mathcal{E}}^{sgs}$ = $\overline{\tau}_{ij}^{sgs} (\partial \overline{u}_i/\partial x_j)$, 
   and residual dissipation, $\mathcal{E}$-RES = $-\overline{\mathcal{E}}^{R}$ = $\tau_{ij}^{R}(\partial \overline{u}_i/\partial x_j)$. At finescale, the plotted  terms are 
   advective transport, T-ADV = $-(1/2) \partial(\overline{u}_j u^{\prime \prime}_i u^{\prime \prime}_i)/\partial x_j$, 
   finescale transport, T-TURB = $-(1/2)\partial(u^{\prime \prime}_j u^{\prime \prime}_i u^{\prime \prime}_i)/\partial x_j$, 
   subgrid dissipation, $\mathcal{E}$-SGS = $-\mathcal{E}^{\prime\prime \, sgs}$ = $\tau^{\prime \prime \, sgs}_{ij} (\partial u^{\prime \prime}_i/\partial x_j)$, 
   residual dissipation, $\mathcal{E}$-RES = $-\mathcal{E}^{\prime\prime \, R}$ = $-\tau^{R}_{ij} (\partial u^{\prime \prime}_i/\partial x_j)$, 
   and the transfer term, $Tr = -u^{\prime\prime}_i u^{\prime\prime}_j (\partial \overline{u}_i/\partial x_j)$, split into contributions from horizontal gradients, $Tr_h$, and vertical gradients, $Tr_v$, of the submesoscale velocity. 
   } 
   \label{fig:ke_budget}
\end{figure}

The dominant terms of submesoscale and finescale KE budgets are frontally-averaged, i.e. over the entire along-front direction and $-1.5 < y < 1.5 \, \rm{km}$ in the cross-front direction, and the vertical profiles of the frontal averages are plotted in Fig.~\ref{fig:ke_budget}. In the following, we first discuss terms in the submesoscale KE balance and then the finescale KE budget.

The submesoscale KE can change due to the buoyancy term, $\overline{B}$, and the dissipation terms, namely, $\overline{\mathcal{E}}, \overline{\mathcal{E}}^{sgs}$, and $\overline{\mathcal{E}}^{R}$. The transport term, $\overline{T}$, represents the transfer of KE between different spatial regions by various processes such as advection, pressure work and the work by various stresses -- molecular, subgrid and residual. The dominant terms of the submesoscale KE budget, Eq.~\ref{eq:les_mean_ke}, are plotted in Fig.~\ref{fig:ke_budget}a as a function of depth at $t = 79.7 \, \rm{h}$. 

Figure~\ref{fig:ke_budget}a shows that the {\em buoyancy production} ($\overline{B}$) is the main source of submesoscale KE in the mixed layer that extends to 50 m depth. $\overline{B}$ represents conversion into KE of the  available potential energy associated with the horizontal buoyancy jump across the front. Below the mixed layer, the buoyancy term is negative. In the bulk of the mixed layer, except for the near-surface region from $z =0$ to approximately $10\, \rm{m}$ depth, the buoyancy production is balanced mainly by the pressure transport, computed as $-(\bar{u}_j/\rho_0)(\partial \bar{p}/\partial x_j)$.
The pressure transport redistributes energy from the 10-40 m depth to both the near-surface region and to the region below 40 m depth. We find that $\overline{B}$ is approximately balanced by $-(\bar{w}/\rho_0) (\partial \bar{ p}/\partial z)$ over the domain. 
This suggests that the overall (horizontally-averaged over the front) submesoscale is in approximate hydrostatic balance for this moderate-strength front.

The \hl{dissipation} terms are also plotted in Fig.~\ref{fig:ke_budget}a. The dissipation due to molecular viscosity is relatively small, and only the contributions from $\overline{\mathcal{E}}^{sgs}$ and $\overline{\mathcal{E}}^{R}$ are shown. The figure shows that both terms are significant in the near surface region, above $10 \, \rm{m}$, and have negative values which indicates removal of submesoscale KE. Note that $\overline{\mathcal{E}}^{sgs}$ is negative everywhere and represents dissipation of submesoscale KE by the LES subgrid model. On the other hand, $\overline{\mathcal{E}}^{R}$ can be either positive or negative and represents exchange of KE between the submesoscale and finescale. The role of $\overline{\mathcal{E}}^{R}$ in this exchange can be readily noticed by expressing $\tau_{ij}^R$ in the Galilean invariant components as suggested by \hl{\citet{Germano:1986}}: $\tau_{ij}^R = \mathcal{L}_{ij}^{o} + \mathcal{C}_{ij}^{o} + \mathcal{R}_{ij}^{o},$  where $\mathcal{L}_{ij}^{o} = \overline{\overline{u}_i \overline{u}_j} - \overline{\overline{u}}_i \overline{\overline{u}}_j$ are the Leonard stresses,  $\mathcal{C}_{ij}^{o} = \overline{\overline{u}_i u^{\prime \prime}_j} + \overline{u^{\prime \prime}_i\overline{u}_j}-\overline{\overline{u}}_i\overline{u^{\prime \prime}}_j-\overline{u^{\prime \prime}}_i \overline{\overline{u}}_j$ are the cross stresses, and $\mathcal{R}_{ij}^{o} = \overline{u^{\prime \prime}_i u^{\prime \prime}_j}-\overline{u^{\prime \prime}}_i \overline{u^{\prime \prime}}_j$ are the stress tensor terms similar to the subgrid Reynolds stresses. Consequently, corresponding to stresses $\mathcal{R}_{ij}^{o} \approx \overline{u^{\prime \prime}_i u^{\prime \prime}_j}$, the term $\overline{Tr} = \overline{u^{\prime \prime}_i u^{\prime \prime}_j} (\partial \overline{u}_i/\partial x_j)$ appears which is similar to the transfer term, $Tr$, in the finescale kinetic energy budget. The frontal average of \hl{$\overline{\mathcal{E}}^{R}$}, however, is negative and indicates a net transfer of energy to the finescale. In the near surface region, transport by the subgrid stresses, $\overline{\tau}_{ij}^{sgs}$, and residual stresses, $\tau_{ij}^{R}$, are also important. The frontal averages of the transports by subgrid and residual stresses are negative and tend to remove the excess energy generated by the significantly large pressure transport in the region.

The dominant terms of the finescale KE budget are plotted in Fig.~\ref{fig:ke_budget}b. The {\em transfer term}, $Tr$, acts as the main source of finescale KE. $Tr$ is split into two parts following \hl{\citet{SullivanM:2018}}: $Tr_h$ with all the terms containing horizontal gradients of the submesoscale velocity and $Tr_v$ with all the terms containing the vertical gradients of the submesoscale velocity. Thus,
\begin{align}
 Tr_h &=  - \left( u_1^{\prime \prime} u_1^{\prime \prime} \frac{\partial \overline{u}_1}{\partial x_1} + 
       u_1^{\prime \prime} u_2^{\prime \prime} \frac{\partial \overline{u}_1}{\partial x_2} +
       u_2^{\prime \prime} u_1^{\prime \prime} \frac{\partial \overline{u}_2}{\partial x_1} +
       u_2^{\prime \prime} u_2^{\prime \prime} \frac{\partial \overline{u}_2}{\partial x_2} +
       u_3^{\prime \prime} u_1^{\prime \prime} \frac{\partial \overline{u}_3}{\partial x_1} + 
       u_3^{\prime \prime} u_2^{\prime \prime} \frac{\partial \overline{u}_3}{\partial x_2} \right), \\ \notag
 Tr_v &=  - \left( u_1^{\prime \prime} u_3^{\prime \prime} \frac{\partial \overline{u}_1}{\partial x_3} + 
        u_2^{\prime \prime} u_3^{\prime \prime} \frac{\partial \overline{u}_2}{\partial x_3} +
        u_3^{\prime \prime} u_3^{\prime \prime} \frac{\partial \overline{u}_3}{\partial x_3} \right).
\end{align}
Figure~\ref{fig:ke_budget}b shows that, in the near-surface region (approximately top 10 m of the mixed layer),  $Tr_h$ is large in magnitude  relative to  $Tr_v$.
However, further below the free surface, $Tr_v$ is the dominant source term. Below the mixed layer, the transfer terms $Tr_h$ and $Tr_v$ are small. \hl{It can also be observed} that below the near-surface region, $Tr_v$ and the subgrid dissipation, $\mathcal{E}^{\prime \prime \, sgs}$ are the dominant terms, and approximately balance each other. 

\begin{figure}[h!]
   \centering
   \includegraphics[clip=true, width=4.0in]{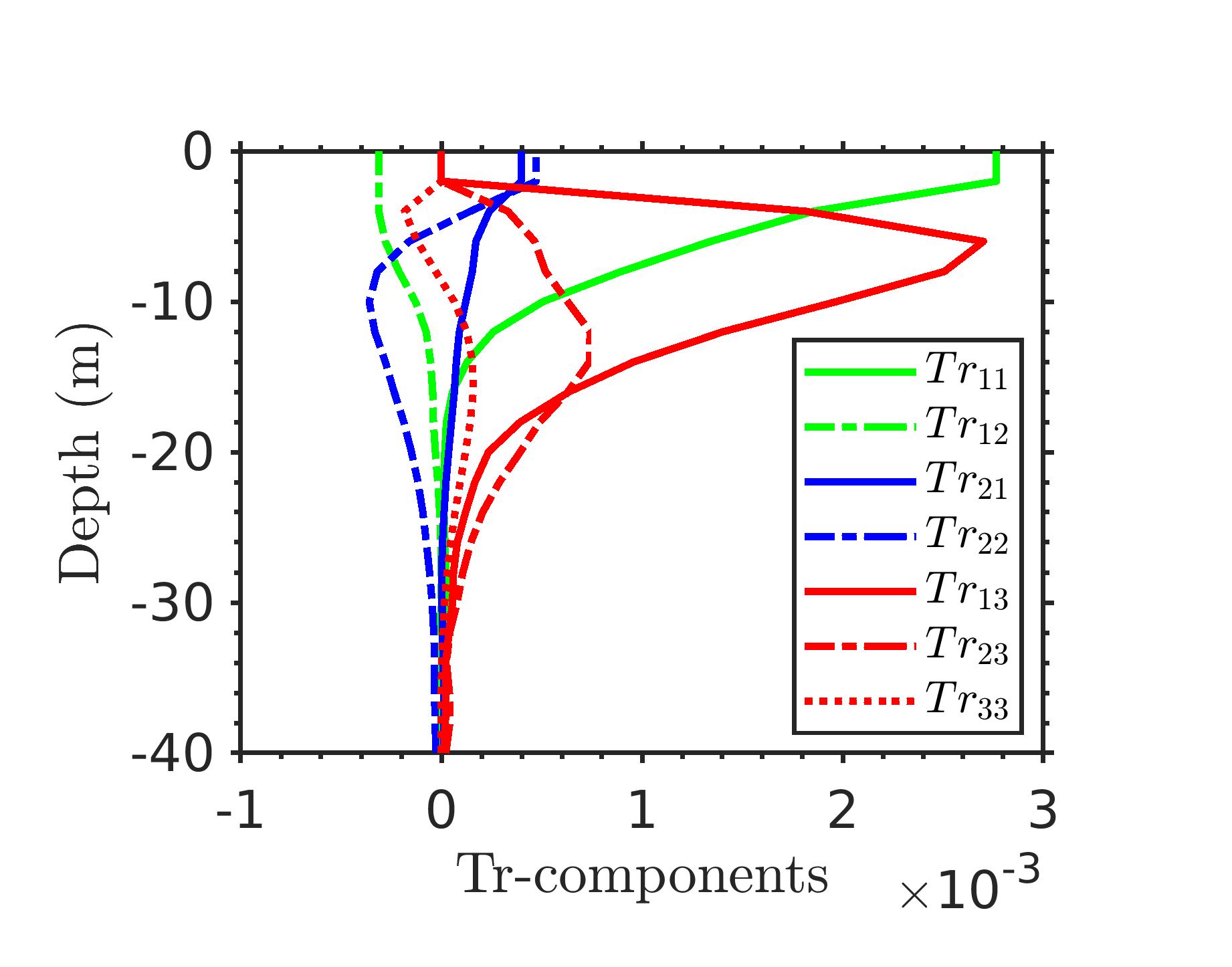} 
   \caption{Dominant components in the transfer term, $Tr$, at $t = 79.7 \, \rm{h}$. Each term has been normalized with $U_0^2f$. Note that the component $Tr_{\alpha \beta}$ denotes $-u_\alpha^{\prime \prime} u_\beta^{\prime \prime} (\partial \overline{u}_\alpha/\partial x_\beta)$ with  no summation over the Greek subscripts, $\alpha$ and $\beta$. $Tr_{31}$ and $Tr_{32}$ are negligible compared to the other terms and not plotted.}  
   \label{fig:Tr_components_n011701}
\end{figure}

The velocity gradient tensor (${\partial \overline{u}_i}/{\partial x_j}$) in the transfer ($Tr$) between the submesoscale and the  finescale has components that can be related to either shear (off-diagonal terms) or strain (diagonal terms) in the flow. The relative importance of the shear and strain components to $Tr$ is further explored in Fig.~\ref{fig:Tr_components_n011701} where front-averaged profiles of each component are plotted as a function of depth. Clearly, $Tr$ is strongly influenced by $Tr_{11} = - u_1^{\prime \prime} u_1^{\prime \prime} ({\partial \overline{u}_1}/{\partial x_1})$ (part of $Tr_h$) and $Tr_{13} = - u_1^{\prime \prime} u_3^{\prime \prime} ({\partial \overline{u}_1}/{\partial x_3})$ (part of $Tr_v$) with the former related to horizontal strain and the latter to vertical shear. In the top 6 m, $Tr_{11}$ \hl{dominates, pointing} to the importance of the  horizontal compressive strain associated with frontogenesis to the enhancement of turbulence. Below 10 m, vertical shear (${\partial \overline{u}_1}/{\partial x_3}$ and additionally ${\partial \overline{u}_2}/{\partial x_3}$) is important. Interestingly, the frontal mean of $Tr_{12} = - u_1^{\prime \prime} u_2^{\prime \prime} ({\partial \overline{u}_1}/{\partial x_2})$ is negative. This reverse energy transfer to horizontal eddying motions in physical space is consistent with the notion that, in this flow, finescale turbulence is organized into the coherent submesoscale in addition to the usual downscale energy transfer to turbulence. The sum of all the components of $Tr$ is positive, i.e., overall, the submesoscale gradients act as a source of the finescale KE. Among gradients of the vertical velocity, only $Tr_{33} = - u_3^{\prime \prime} u_3^{\prime \prime} ({\partial \overline{u}_3}/{\partial x_3})$ involving vertical strain is significant; however it is smaller than the dominant term $Tr_{13}$. 

In addition to $Tr$, $\mathcal{E}^{\prime \prime \, R}$ is another dominant source of the finescale kinetic energy, mainly near the surface.  $\mathcal{E}^{\prime \prime \,R}$ represents  the interaction of the residual stress  with the finescale shear. There is only a partial balance between the production of the kinetic energy and dissipation by the subgrid stresses, $\mathcal{E}^{\prime \prime \, sgs}$. In this region, transport by both the submesoscale and finescale velocity fields is significant and removes energy from the layer.

\begin{figure}[h!]  
\centering
\includegraphics[clip=true, width=6.0in]{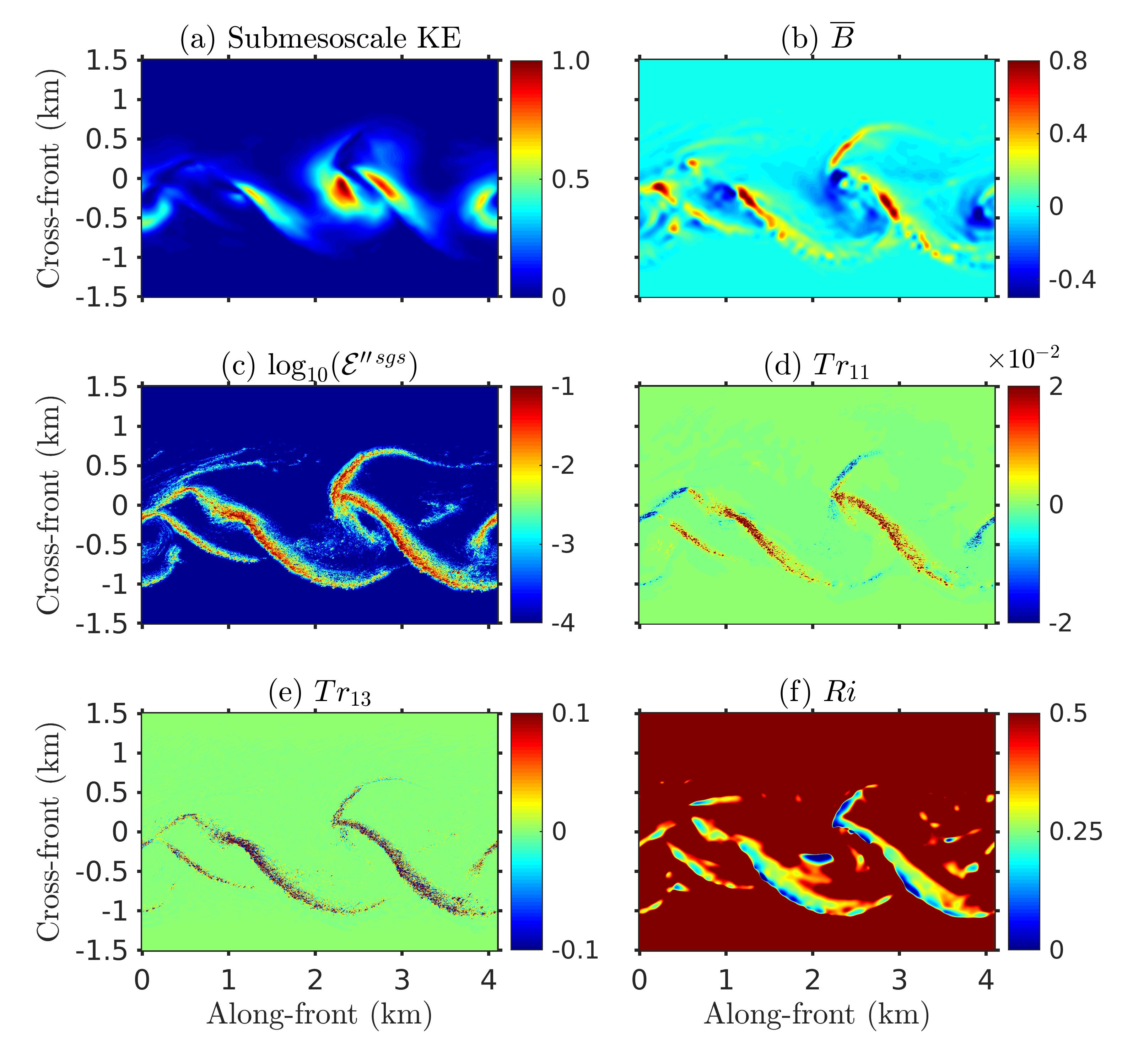}
  \caption{Variation at 10 m depth and $t = 79.7 \, \rm{h}$ of submesoscale fluctuation (downfront mean removed) kinetic energy (a),  selected energy budget terms at the submesoscale and the finescale (b-e),  and submesoscale gradient Richardson number, $Ri$ (f). Here, $Ri = (\partial \bar{b} /\partial z)[(\partial \bar{u}/\partial z)^2 + (\partial \bar{v}/\partial z)^2]^{-1}$. The submesoscale kinetic energy is normalized with $U_0^2/2$ and the energy budget terms are normalized with $U_0^2 f$.  
 }\label{fig:ke_budget_contours}
\end{figure}

The horizontal organization of \hl{selected terms} that appear in the kinetic energy balance are shown in Fig.~\ref{fig:ke_budget_contours} at 10 m depth. The local value of submesoscale buoyancy production, $\overline{B}$ (Fig.~\ref{fig:ke_budget_contours}b) has both positive  regions (primarily in the filaments) and negative  regions (primarily in the eddy cores).  However, the horizontal average of $\overline{B}$  is positive (Fig.~\ref{fig:ke_budget}a) and indicates restratification accompanied by the conversion of potential energy in the lateral stratification to submesoscale KE. The resemblance between the horizontal structures of submesoscale KE and buoyancy production is apparent in Figs.~\ref{fig:ke_budget_contours}(a,b), further reinforcing the importance of  $\overline{B}$ for the submesoscale KE. The dissipation of the finescale (Fig.~\ref{fig:ke_budget_contours}c) is concentrated in the vortex filaments, again  pointing to the spatially localized, coherent organization of  dissipative turbulence in this problem. The dominant terms that transfer energy from  the submesoscale to the finescale owing to horizontal strain (Fig.~\ref{fig:ke_budget_contours}d) and vertical shear (Fig.~\ref{fig:ke_budget_contours}e) are also localized in the vortex filaments.  The gradient Richardson number ($Ri_g$)  based on vertical shear and stratification of the submesoscale is plotted in Fig.~\ref{fig:ke_budget_contours}f. The regions with subcritical $Ri_g < 0.25$ are within those where $Tr_{13}$ is large. Thus, local shear instability is a driver of turbulence at these locations. 

\section{Frontogenesis} \label{sec:frontogenesis}
The formation of coherent structures, i.e., the vortex filaments and the coherent submesoscale eddies, show frontogenesis, notably large increase in $M^2/f^2$ from its initial value. The role of the coherent structures in vertical transport was addressed in  section  \ref{sec:scale_separation}. In this section, the focus is on the processes that are responsible for frontogenesis and those which counteract to balance it.  We investigate the time rate-of-change of horizontal buoyancy gradient at the submesoscale, $\boldsymbol{\nabla_h} \overline{b}$, where $\boldsymbol{\nabla_h} = \partial_x \hat{\mathbf{i}} + \partial_y \hat{\mathbf{j}}$ and $\hat{\mathbf{i}}$ and $\hat{\mathbf{j}}$ are unit vectors in x- and y-directions, respectively (e.g., \citet{CapetMMS:2008b}). It can be noted that buoyancy is directly related to the temperature deviation, $b = g\alpha T$. In particular, the following dynamical equation is quantified:
\begin{align} \label{eq:frontogenesis}
 \frac{1}{2} \frac{D}{Dt} |\boldsymbol{\nabla_h} \overline{b}|^2 &= \boldsymbol{\nabla_h} \overline{b} . (\mathbf{Q_s} + \mathbf{Q_w} + \mathbf{Q_{dh}} + \mathbf{Q_{dv}}), \notag \\ 
 & = F_s + F_w + F_{dh} + F_{dv}.
\end{align}
In the above equation, $\mathbf{Q_s}$ and $\mathbf{Q_w}$ denote contributions due to advection by the submesoscale and are given  by 
\begin{align}
 \mathbf{Q_s} &= -(\partial_x \overline{u} \, \partial_x \overline{b} + \partial_x \overline{v} \, \partial_y \overline{b} ,  \; \partial_y \overline{u} \, \partial_x \overline{b} + \partial_y \overline{v}  \, \partial_y \overline{b}), \notag \\
 \mathbf{Q_w} &= -(\partial_x \overline{w} \, \partial_z \overline{b}, \; \partial_y \overline{w} \, \partial_z \overline{b}).
\end{align}
Thus, $\mathbf{Q_{s}}$ \hl{corresponds to the straining} of the buoyancy field by horizontal motion and $\mathbf{Q_{w}}$ is an analogous term related to vertical motion. The contributions from diabatic processes -- \hl{molecular diffusion, and subgrid and residual fluxes} -- are included in $\mathbf{Q_{dh}}$ and $\mathbf{Q_{dv}}$. Between the two, $\mathbf{Q_{dh}}$ is associated with the horizontal diabatic processes, and $\mathbf{Q_{dv}}$ is associated with the vertical diabatic processes, 
\begin{align}
 \mathbf{Q_{dh}} &= \boldsymbol{\nabla_h} (\boldsymbol{\nabla_h}.(\kappa \boldsymbol{\nabla_h} \overline{b}) - g \alpha \boldsymbol{\nabla_h}.(\mathbf{\overline{q}_h^{sgs}} + \mathbf{q_h^R})), \notag \\
 \mathbf{Q_{dv}} &= \boldsymbol{\nabla_h} ( \partial_z (\kappa \partial_z \overline{b}) - g \alpha \partial_z (\overline{q}_v^{sgs}+q_v^{R})).
\end{align}
The forcing terms on the right hand side of Eq.~\ref{eq:frontogenesis}, $F_i$, where $i = \{s, w, dh, dv\}$, are obtained by taking the dot-product of $\boldsymbol{\nabla_h} \overline{b}$ with appropriate $\mathbf{Q}$-terms.

\begin{figure*}[h!]
   \centering
   \includegraphics[clip=true, width=6.0in]{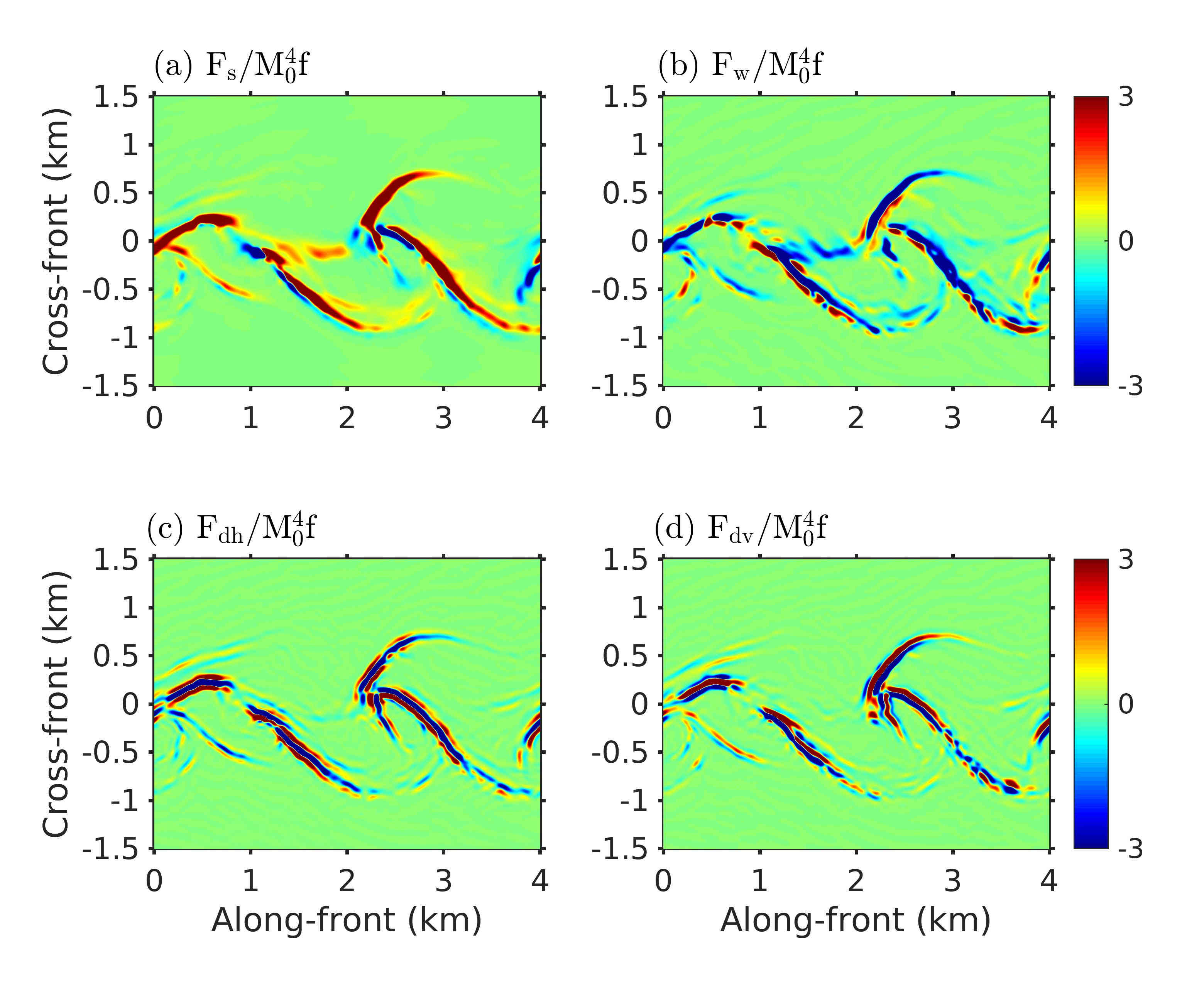} 
   \caption{Different terms in the RHS of the transport equation for 
   $|\boldsymbol{\nabla_h} \overline{b}|^2$ (Eq.~\ref{eq:frontogenesis}) are shown at $10 \, \rm{m}$ depth and $t = 79.7 \, \rm{h}$.}
   \label{fig:frontogenesis_depth10}
\end{figure*}

\begin{figure*}[h!]
   \centering
   \includegraphics[clip=true, width=6.0in]{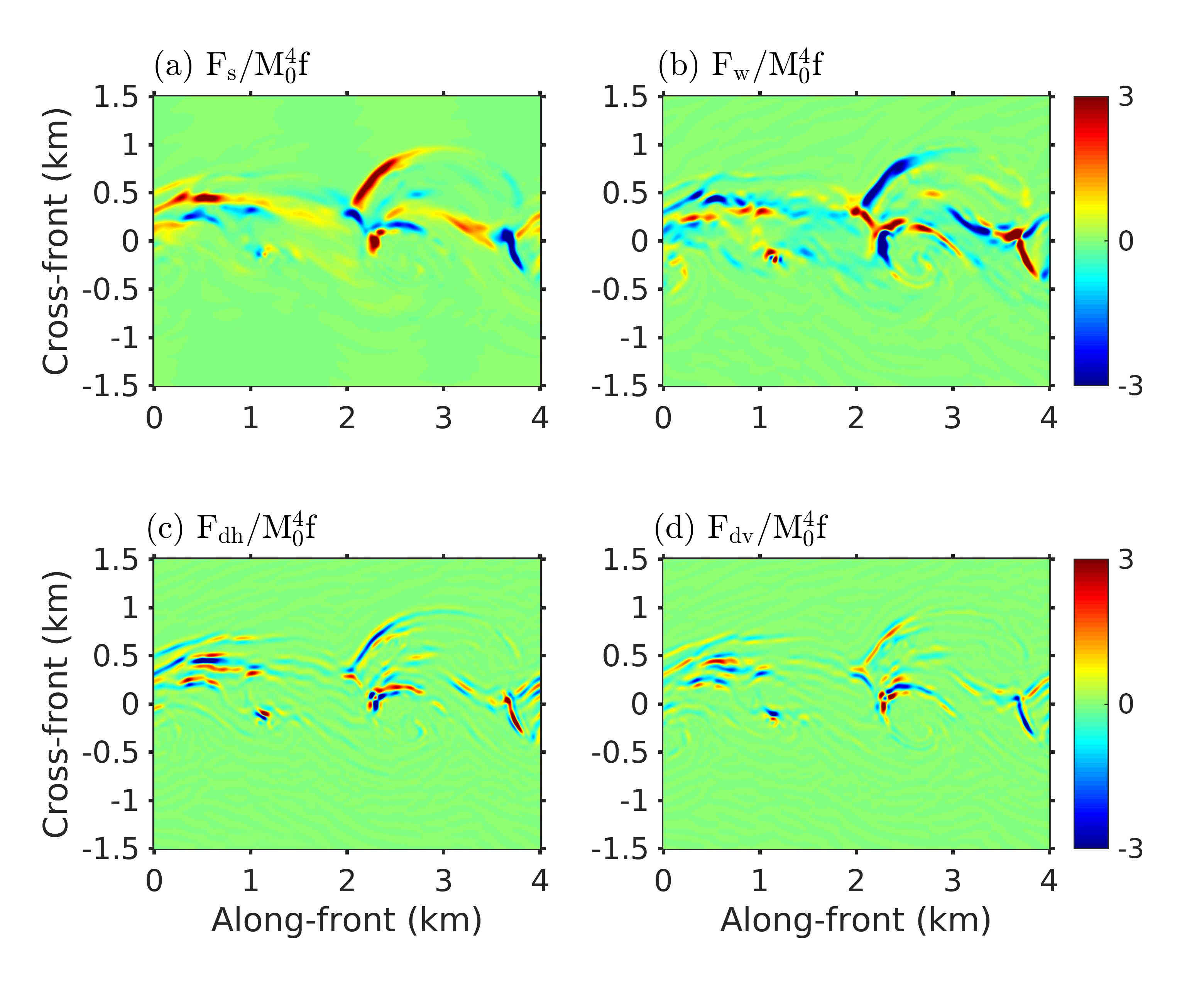} 
   \caption{Different terms in the RHS of the transport equation for 
   $|\boldsymbol{\nabla_h} \overline{b}|^2$ (Eq.~\ref{eq:frontogenesis}) are shown at $30 \, \rm{m}$ depth and $t = 79.7 \, \rm{h}$.} 
   \label{fig:frontogenesis_depth30}
\end{figure*}

\begin{figure}[h!]
   \centering
   \includegraphics[clip=true, width=4.0in]{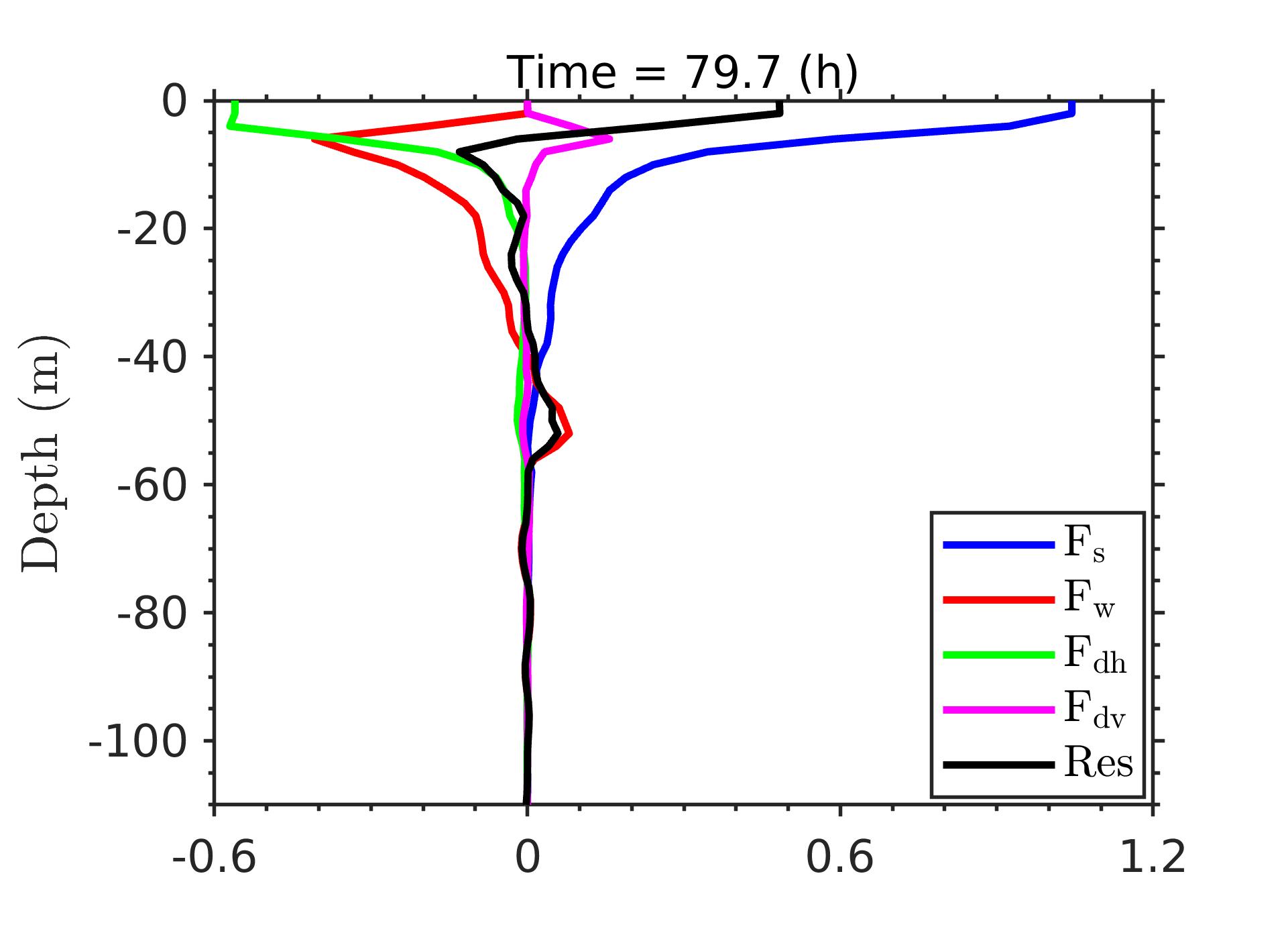} 
   \caption{Forcing terms in Eq.~\ref{eq:frontogenesis} at t = 79.7 h. Each term is normalized with $M_0^4 f$. Here, the residual is defined by $\rm{Res} = F_s + F_w + F_{dh} + F_{dv}$.} 
   \label{fig:frontogenesis_vertical_profile}
\end{figure}

In Figs.~\ref{fig:frontogenesis_depth10} and \ref{fig:frontogenesis_depth30}, the forcing terms are plotted at depths $10 \,\rm{m}$ and $30 \, \rm{m}$, respectively. Near the surface, $F_s$ is predominantly positive in the vortex filaments and is frontogenetic, whereas $F_w$ is predominantly negative and is frontolytic. On the other hand, $F_{dh}$ and $F_{dv}$ are both frontogenetic and frontolytic in these regions -- $F_{dh}$ being frontolytic in the center and frontogenetic at the edges, while $F_{dv}$ being frontogenetic at the center and frontolytic at the edges. Near the mid-depth at $30 \, \rm{m}$, the forcing terms $F_{s}$ and $F_{w}$ have both positive and negative values, relatively more interspersed at the front. The magnitudes of the forcing terms $F_s$ and $F_w$, as well as $F_{dh}$ and $F_{dv}$, are much smaller compared to those near the surface. A qualitative balance between $F_s$ and $F_w$ and that between $F_{dh}$ and $F_{dv}$ can be observed at both depths. The vertical profiles of frontal-averaged ($0 < x < L_x$ and $-1.2 < y < 1.2 \, \rm{km}$) forcing terms are shown in Fig.~\ref{fig:frontogenesis_vertical_profile}. Three distinct regions can be identified -- upper, intermediate and bottom regions. In the upper, near-surface region (above $10 \, \rm{m}$ depth), $F_s$ is balanced by both $F_w$ and $F_{dh}$. However, very close to the surface $F_{w}$ goes to zero and the balance is provided only by $F_{dh}$. The balance between $F_s$ and $F_{dh}$ is partial at the surface, i.e. the frontal-averaged value of $|\boldsymbol{\nabla_h} \overline{b}|^2 $ continues to increase with time because the length of the thin O(100) m wide frontal regions continues to increase as BI proceeds. 

In the intermediate region between $10-40 \, \rm{m}$ depth, there is \hl{a dominant balance} between $F_s$ and $F_w$; $F_s$ is predominantly frontogenetic and $F_w$ is predominantly frontolytic. This can also be observed in Figs.~\ref{fig:frontogenesis_depth10} and \ref{fig:frontogenesis_depth30}. Finally, near the bottom between $40-60 \, \rm{m}$ depth, there is relatively weak frontogenesis that arises from $F_w$. In all the three regions, the frontal averaged $F_{dv}$ is small.  

The balance of the frontal-averaged buoyancy gradient described above corresponds to $t = 79.7 \, \rm{h}$ and reflects the behavior during the time the coherent eddies grow. Before the growth of BI, the forcing terms, $\langle F_i \rangle_{xy}$, remain small. As BI starts growing at $t \approx 40 \, \rm{h}$, $F_i$ begin to develop near the surface. When the eddies form and grow, $\langle F_i \rangle_{xy}$ become significantly large (Fig.~\ref{fig:frontogenesis_vertical_profile}). During this period the submesoscale KE also grows. Afterwards when the submesoscale KE saturates at $t \approx 100 \, \rm{h}$, $\langle F_i \rangle_{xy}$ remain significant only in the upper region. In the intermediate and bottom regions, the magnitudes of frontal averaged forcing terms become small. 
  
\section{Discussion and conclusions} \label{sec:discussion_and_conclusion}
The spin-down of a mixed layer front is studied using a large eddy simulation (LES) model.  The front of width 1.2 km is initially in geostrophic balance,  has a moderate Rossby number of 0.32, and  is confined to a weakly stratified ($Ri = 0.26$) mixed layer of thickness 50 m that lies above a strongly stratified thermocline.  The non-hydrostatic, Boussinesq Navier-Stokes equations are  numerically solved on a turbulence-resolving isotropic grid that discretizes a domain that is $4098 \, \rm{m}$ in the along-front, $6146 \, \rm{m}$ in the cross-front, and $130 \, \rm{m}$ in the vertical with resolution of $2\,\rm{m}$ in all three coordinate directions. The high-resolution LES enables us to simulate submesoscale eddies,  three-dimensional turbulence and their interaction. 

The initial configuration with $Ri = 0.26$ is such that the potential vorticity is negative over the front, making it unstable to symmetric instability (SI)  that then develops secondary KH instability~\citep{TaylorF:2009}  and three-dimensional turbulence~\citep{AroboneS:2015}.  Although SI grows initially, the baroclinic instability (BI) soon becomes the dominant mode as $Ri$ increases to beyond 0.95 during restratification of the front~\citep{HaineM:1998,FoxKemperFH:2008}. The growth of BI is frontogenetic \citep{Hoskins:1982} and results in the formation of thin vortex filaments at the front. Furthermore, long vortex filaments, comparable in length to that of the wavelength of the dominant BI mode, roll up to produce coherent submesoscale eddies. These coherent structures are localized in space and  characterized by large values of strain rate, vorticity, and density gradient. The local Rossby number in these structures is $O(10)$ indicating local loss of rotational control. 

Previous studies of the mixed layer instability  with turbulence-resolving \hl{simulations (e.g. \citet{SkyllingstadS:2012,StamperT:2017})} have found vortex filaments, submesoscale eddies and turbulence during the flow evolution. 
In the present simulation of an isolated, unforced front, the turbulence that develops during BI occurs in localized regions at the sharpening filament structure similar to \citet{SkyllingstadS:2012} who simulated the evolution of \hl{a warm filament} (double-front configuration with lateral periodicity). Given the sparseness and locality in physical space of the turbulence that develops in this problem, we are  motivated to examine scale interactions in physical space rather than the classical wavenumber-based approach employed by  \citet{SkyllingstadS:2012}. Therefore, unlike previous studies of the mixed layer instability, we decompose the flow into a submesoscale field and a finescale field in physical space, describe the properties of these two fields, and quantify their interaction. \hl{The classical decomposition into an overall frontal average and a fluctuation would obscure the coherent submesoscale structures and prevent us from elucidating their important role in frontal dynamics. The present decomposition enables the direct quantification of how the coherent submesoscale  contributes to processes such as transfer of mass, momentum and buoyancy to the bottom of the mixed layer and the interaction between the mixed layer and the pycnocline.} A low-pass Lanczos spatial filter is used to decompose a given flow variable $\phi$ into the sum of a submesoscale component $\overline{\phi}$ and a finescale component $\phi^{\prime \prime}$. Our finding of a change of kinetic energy spectrum slope  and flow anisotropy at O (100) m length scale guides the choice of filter scale as O(100) m. 

\begin{figure*}[h!]
 \centering
 \includegraphics[clip=true, width=6.0in]{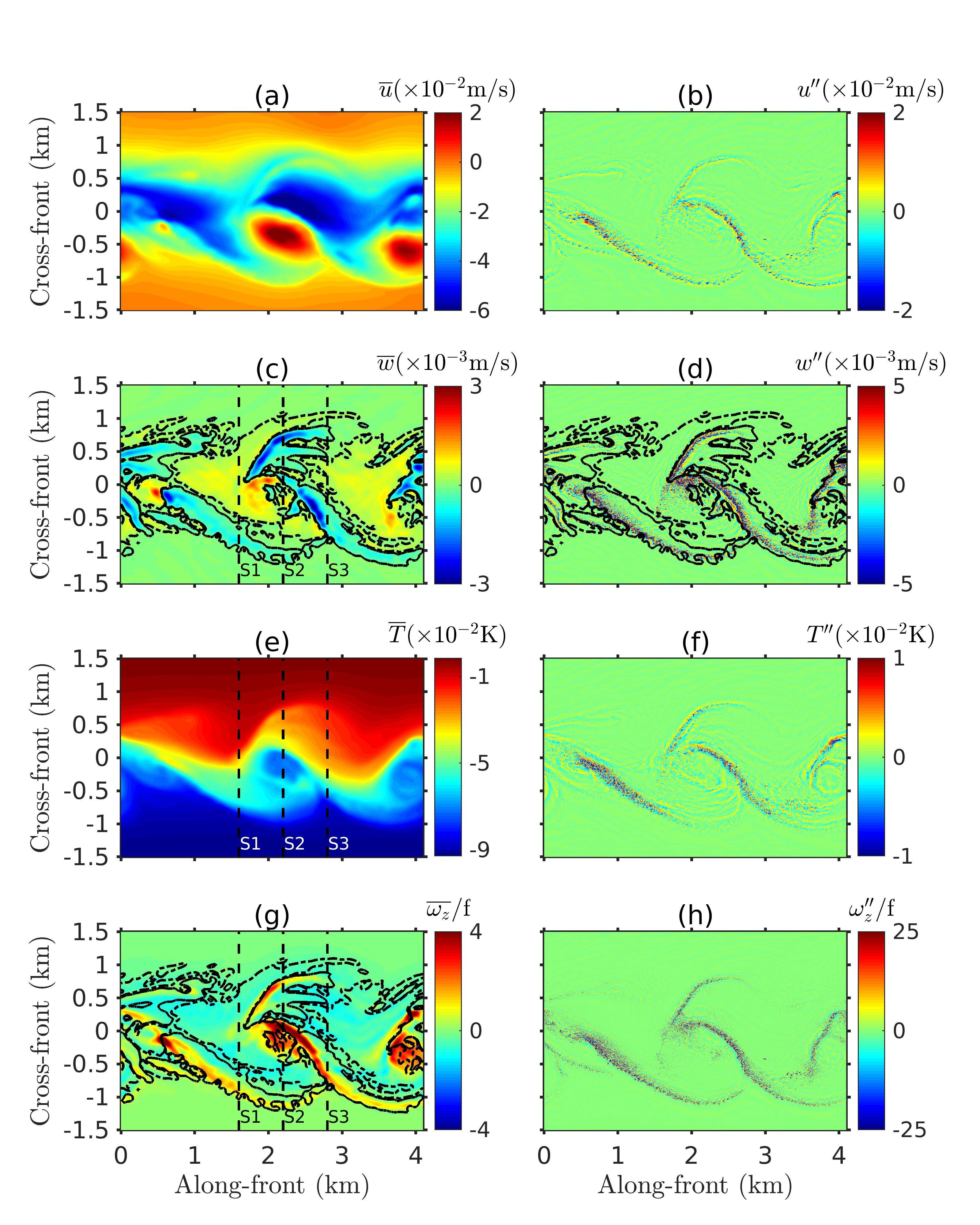}
 \caption{Submesoscale (left column) and finescale (right column) fields plotted at $10\,\rm{m}$ depth at $t = 86\, \rm{h}$: (a, b) along-front velocity, (c, d) vertical velocity, (e, f) temperature, and (g, h) non-dimensional vertical vorticity. The solid black lines and the dash-dot black lines in panels (c, d, g) correspond to $\overline{w} = - 0.1 \,\rm{mm\,s^{-1}}$ and $\overline{w} = 0.1 \,\rm{mm\,s^{-1}}$, respectively, and approximately enclose the regions with downwelling and upwelling motions. Three vertical cross-sections S1, S2 and S3 at $x = 1.6, 2.2$ and $2.8 \, \rm{km}$, respectively, are also shown in panels (c, e, g).}
 \label{fig:scale_separation}
\end{figure*}

\begin{figure*}[h!]
   \centering
   \includegraphics[clip=true, width = 6.0in]{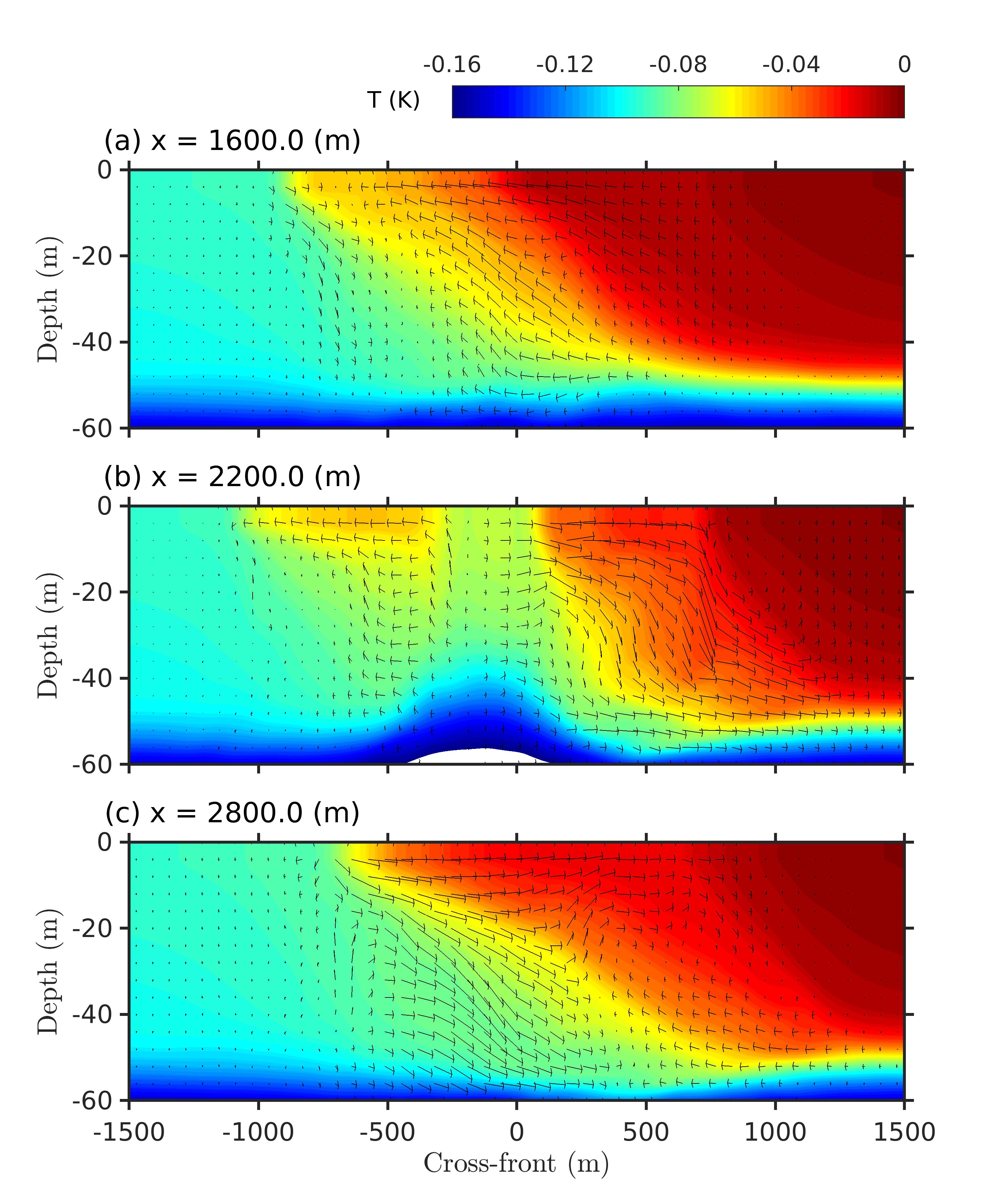} 
   \caption{Ageostrophic secondary circulation at three  different along-front locations (S1, S2 and S3) that were marked in Fig.~\ref{fig:scale_separation}(e). The secondary circulation is shown by vertical $y-z$ cuts: (a) at  S1 ($x = 1.6 \, \rm{km}$), (b) at S2  ($x = 2.2 \, \rm{km}$),  and (c) at S3 ($x = 2.8 \, \rm{km}$). The color contours show the temperature, and the arrows show the velocity vectors in the vertical planes.}  
   \label{fig:BI_SAC}
\end{figure*}

We find that the submesoscale component (e.g. left column of Fig.~\ref{fig:scale_separation}) represents the coherent structures that emerge during the evolution of the mixed layer instability and are distributed throughout the frontal region. The finescale component (e.g. right column of  Fig.~\ref{fig:scale_separation}) represents small-scale turbulence that is {\em spatially organized}  and  concentrated in filamentary regions where the submesoscale field has large spatial gradients. 

An important feature of submesoscale dynamics in frontal regions is the associated increase in vertical transport. The present simulation shows that $w$ exceeds $0.005\, \rm{m\,s^{-1}}$ in several regions and becomes as large as $0.02 \, \rm{m\,s^{-1}}$ within vortex filaments. Regions of positive and negative $w$ are interspersed; it is the decomposition of $w$ into the submesoscale ($\bar{w}$) and the finescale  ($w^{\prime \prime}$) that allows us to separate upwelling from downwelling regions. Downwelling by  $\bar{w}$ is concentrated at the filaments that curve into the submesoscale vortices while upwelling occurs over more spatially extensive regions in and around these vortices. The associated buoyancy flux is restratifying.  Although the instantaneous magnitude of $w^{\prime \prime}$ has an order of magnitude large peak relative to $\bar{w}$, the additional net contribution by the  finescale to the overall transport is small. 

An examination of the submesoscale  velocity over $y-z$ cross sections at different $x$ (along-front) locations shows significant along-front variability. Thus, the secondary ageostrophic circulation, taken to be two-dimensional in simplified along-front averaged models, is three-dimensional owing to the action of the filaments and vortices in the flow.  We define back-to-forward direction as aligned with the  jet velocity (negative $x$-direction). Heavier fluid with downwelling velocity is found in the filament structure in the back regions of the eddies while the lighter fluid with upwelling velocity is found in the forward regions. 

The coherent structures have enhanced relative vorticity. Cold filaments with denser fluid have concentrated positive submesoscale $\overline{\omega_z}$ that roll up into cyclonic submesoscale vortices with vorticity as large as $\omega_z/f \approx 5$ in the surface layer similar to the simulations of \citet{GulaMM:2015} with 150 m horizontal resolution. The finescale vorticity, ${\omega^{\prime\prime}_z}$, is an order of magnitude larger and is concentrated in the thin filaments. 

The finescale kinetic energy (KE) is found to increase rapidly during the initial 20 h when SI is active and then its frontal average saturates although the local spatio-temporal variability is substantial. The submesoscale KE continues to increase after 20 h when SI becomes nonlinear; however, the bulk of the increase occurs during  BI. The turbulent finescale is dissipative and an outstanding question is what maintains the energy of the finescale. To answer this question and have a better understanding of the energy pathways to and between the two components, we evaluate the KE transport equations separately for the submesoscale and finescale, and report vertical profiles of horizontally-averaged terms at a late time ($t$ = 79.7 h) during the BI evolution. 

With regards to the submesoscale KE, the buoyancy production term, which is significant between 10 m and the mixed layer base, acts as the main source. The pressure-transport term  redistributes KE spatially from lower in the mixed layer to the 10 m near-surface region. Both the residual stress (explicitly computed from the flow field as $\tau_{ij}^{R} = \overline{u_i u_j} - \overline{u}_i \overline{u}_j$) and the subgrid stress (modeled)  act to dissipate submesoscale KE. The dissipation by residual stress, $\overline{\mathcal{E}}^{R}$, is larger than the dissipation by subgrid stress, $\overline{\mathcal{E}}^{sgs}$, by a factor of about 2. Direct molecular dissipation is negligible at the O(100) m scale of the submesoscale.

\hl{The finescale KE is found} to be primarily delivered by the transfer term, $ Tr = -u^{\prime \prime}_i u^{\prime \prime}_j (\partial \overline{u}_i/\partial x_j)$, which represents the interaction of the finescale with the submesoscale velocity gradients. The transfer term, $Tr$, is further decomposed into ${Tr}_h$ \hl{(which involves horizontal gradients)} and $ {Tr_v}$. As in several of the other diagnostics, the near-surface upper 10 m layer is qualitatively different in that $Tr_h$ ($Tr_{11}$) is the primary source of finescale KE in contrast to the remainder of the mixed layer where it is $Tr_v$ ($Tr_{13}$) that dominates. The buoyancy flux has a negligible contribution to the KE balance. This is in contrast to the submesoscale, where it is the buoyancy flux which energizes the submesoscale KE. The production of finescale KE by the residual stress $\tau_{ij}^{R}$ is an important contributor to the budget only in the top 10 m.

In the present study, horizontal strain ($\partial u/\partial x$) and vertical shear ($\partial u/\partial z$) of the submesoscale velocity field act as dominant production terms of finescale kinetic energy at different depths. In particular, in the 6 m near-surface region, it is the horizontal strain which dominates the transfer while, in the remainder of the mixed layer, the transfer is mediated by the vertical shear. \citet{SullivanM:2018} simulated the evolution of a cold (dense) filament in a  background of strong ocean boundary layer (OBL) turbulence. BI did not emerge in the mixed layer, but there was frontogenesis by the ageostrophic circulation instigated by OBL turbulence leading to frontal turbulence. They found that in the down-front wind case and no wind (surface cooling) case, the horizontal shear was the dominant production term for the TKE while in the cross-front wind case it was the vertical shear. Thus, there are qualitative differences in frontal turbulence generation between a front undergoing BI with subsequent submesoscale meanders and vortices as in the present case, and a front that develops in a pre-existing field of strong OBL turbulence without exhibiting BI.

During the evolution of BI, horizontal currents ($u$, $v$) develop strong variability in the  horizontal ($x$, $y$) that locally sharpen the front and intensify  the buoyancy gradient ($\mathbf{\nabla_h} b$). The terms in the balance equation for the square of $\mathbf{\nabla_h} b$ are diagnosed to better understand the frontogenetic mechanisms. Instantaneous visualizations at 10 m depth shows active frontogenesis by $F_s$ \hl{(which involves horizontal gradients of horizontal velocity)}, mainly at the filaments, while $F_w$ \hl{(which involves horizontal gradients of vertical velocity)} acts to counteract frontogenesis. \hl{Instantaneously}, the diabatic (diffusive) terms act to both increase and decrease the local buoyancy gradient. An overall picture of the diabatic term emerges after performing horizontal averages in the frontal region. Horizontal diffusion, both resolved and subgrid, is the primary process in the upper 10 m that counteracts frontogenesis by $F_s$. In the remainder of the mixed layer it is primarily $F_w$ (which involves horizontal gradients of vertical velocity) that counteracts frontogenesis by $F_s$. \citet{CapetMMS:2008b} in their analysis of frontogenesis during mesoscale-submesoscale transition in a simulation with 750 m horizontal resolution, found that $F_{dv}$ (vertical diffusion) modeled by k-profile parametrization (KPP) \hl{was the primary term in the model counteracting $F_s$}. Horizontal diffusion ($F_{dh}$) could not be calculated separately because it was implicit in the horizontal advection and  the horizontal grid resolution was insufficient. In the present 2 m resolution LES, the front thins to O(100) m. We find that the contribution from horizontal diffusion ($F_{dh}$), calculated explicitly here, counteracts frontogenesis by $F_s$ in the near-surface layer and, thus, plays an important role in limiting the width of the frontogenetic region.

\hl{In cold filaments that evolve in a turbulent boundary layer, surface frontogenesis results from the  secondary circulation that occurs under the turbulent thermal wind (TTW) condition~\citep{McWilliamsGMRS:2015,SullivanM:2018}. Frontal sharpening proceeded to the grid scale of O (10) m in the 2-D numerical simulations \citep{McWilliamsGMRS:2015} of this problem that were conducted with the  k-profile parametrization (KPP) of vertical mixing,  while \citet{SullivanM:2018} found that frontogenesis was arrested at O(100) m by shear-driven turbulence in a simulation that resolved the turbulent boundary layer. BI is excluded in the 2-D model of \citet{McWilliamsGMRS:2015} whereas, in the model of \citet{SullivanM:2018}, the filament evolution process is fast and BI remains unimportant during this period. Although different from the BI-driven frontogenesis and localized turbulence of the present study, the arrest scale of frontogenesis observed by \citet{SullivanM:2018} is close to the value of O(100) m obtained in the present study. It is worth noting that features sharper than O(100) m can form in fronts. For example, \citet{PhamS:2018} in their study of a strong front with large $M^2/f^2$ find sharpening down to O(10) m that precedes the release of a gravity current.}

\hl{Although the submesoscale dynamics is studied here for a  model front without any external influences, it exhibits processes which are also observed in the real ocean. The study of \citet{DAsaro_etal:2018} using satellite-tracked surface drifters in the northern Gulf of Mexico revealed sharpened fronts which wrap into cyclonic eddies, a process that is resembled by the  rolling-up of the filament structures into cyclonic eddies in the present study. The {\em zipper} structures, which form when two filaments combine into one as they wrap into the eddy, are similar to the joining of filaments structures from opposite sides of the front seen here. Moreover, the  positive (cyclonic) vertical vorticity and downwelling at the front as well as their enhanced  magnitudes in the convergent zipper region found by  \cite{DAsaro_etal:2018} are features that occur in the submesoscale dynamics of the present study. Similar to the asymmetry in the distribution of the positive and negative vertical vorticity observed in the study of \citet{ShcherbinaDLKMM:2013}, we find in the simulation that the anticyconic vertical vorticity is weaker ($\omega_z \approx -1.2f$) and spatially more extensive than the cyclonic vertical vorticity ($\omega_z \approx 5f$) that is concentrated in thin filaments. Furthermore, the large strain rate and positive vertical vorticity at the submesoscale are correlated in the simulation; both of these properties are characteristic of the filaments.} 

The present study underlines the importance of coherent structures, i.e. vortex filaments and coherent submesoscale eddies, in  the dynamics of the spindown of an unforced front in a mixed layer by elucidating their role in vertical transport, frontogenesis and maintenance of fine-scale turbulence. The analysis was based on Eulerian statistics and, in future work, we hope to shed further light on subduction and mixing by Lagrangian tracer analysis.  Surface forcing by wind, buoyancy and waves introduces additional complexities into the evolution of the baroclinic instability  as has been shown by several recent numerical studies, \hl{e.g., \cite{MahadevanTF:2010, HamlingtonRF:2014, WhittT:2017, CalliesF:2018}: cooling at the surface can compete with the restratifying buoyancy flux generated by  BI; a downfront wind can create negative buoyancy flux, counter restratification, and inhibit BI; and, Langmuir turbulence generated by the interaction between the Stokes drift by surface waves and the  wind can inject kinetic energy directly at small scales and enhance mixing at the front.  In several numerical models, BI is found to grow, albeit modified by the forcing and, somewhat surprisingly, even after the submesoscale is exposed to a storm in the study by \cite{WhittT:2017}. As discussed previously in this section, observational studies reveal submesoscale features and dynamics similar to the findings in the present unforced simulation.} The net effect of the external forcing on submesoscale processes is dependent on the specifics of the problem and the strength of the forcing. It will be of interest to systematically study how the submesoscale/finescale properties and their mutual interaction change when other processes influence  the mixed layer instability.

\vspace{5mm}
\noindent \textbf{Acknowledgments}

We are pleased to acknowledge the support of ONR grants N00014-15-1-2578 and N00014-18-1-2137.



 \bibliographystyle{model2-names}\biboptions{authoryear}
 \bibliography{references}


%
%
%

\pagebreak
\begin{table}
  \caption{Influence of the cutoff wavenumber, $k_c$, on the kinetic energy ($E^>$) in large wavenumbers with  $k>k_c$ (i.e., length scales less than or equal to $2\pi/k_c$). Here, $E^>  = \int_{k_c}^{k_{max}} E({k}) d{k}$ where $E(k)$ is the power spectrum of the kinetic energy. $E^>_u$, $E^>_v$ and $E^>_w$ are the contributions to $E^> $ from the velocity components $u$, $v$ and $w$, respectively.}
  \label{tab:energy_content}
 \begin{tabular}{ c c c c c}
  $k_c \, (\rm{rad \, m^{-1}})$ & $E^> \, (\times 10^{-6} \, \rm{m^2s^{-2}})$ & $E^>_u \,(\%)$ & $E^>_v \,(\%)$ & $E^>_w \,(\%)$ \\ \hline \hline
  0.01 & 5.18 & 38 & 51 & 11 \\
  0.02 & 2.79 & 37 & 45 & 18 \\
  0.04 & 1.95 & 35 & 43 & 22 \\
  0.06 & 1.56 & 33 & 43 & 24 \\
  0.08 & 1.30 & 32 & 43 & 25 \\
  0.1  & 1.11 & 30 & 44 & 26 \\ \hline
  \end{tabular}
\end{table}

\end{document}